\documentclass[iop,onecolumn]{emulateapj}
\shortauthors{R. FUSCO-FEMIANO \& A. LAPI}
\shorttitle{ENTROPY FLATTENING, GAS CLUMPING, AND TURBULENCE}

\begin{document}

\title{Entropy Flattening, Gas Clumping and Turbulence\\ in Galaxy Clusters}
\author{R. Fusco-Femiano$^{1}$ and, A. Lapi$^{2,3}$}
\affil{$^1$IAPS-INAF, Via Fosso del Cavaliere, 00133 Roma,
Italy - roberto.fuscofemiano@iaps.inaf.it}\affil{$^2$Dip. Fisica, Univ. `Tor Vergata', Via Ricerca Scientifica
1, 00133 Roma, Italy.}\affil{$^3$SISSA, Via Bonomea 265, 34136 Trieste, Italy.}

\begin{abstract}
Several physical processes and formation events are expected in cluster
outskirts, a vast region up to now essentially not covered by observations.
The recent \textsl{Suzaku} (X-ray) and \textsl{Planck} (Sunayev-Zeldovich
effect) observations out to the virial radius have highlighted in these
peripheral regions a rather sharp decline of the intracluster gas
temperature, an entropy flattening in contrast with the theoretically
expected power law increase, the break of the hydrostatic equilibrium even
in some relaxed clusters, a derived gas mass fraction above the cosmic
value measured from several CMB experiments, and a total X-ray mass lower
than the weak lensing mass determinations. Here we present the analysis of
four clusters (A1795, A2029, A2204 and A133) with the SuperModel that
includes a nonthermal pressure component due to turbulence to sustain the
hydrostatic equilibrium also in the cluster outskirts. In such way we
obtain a correct determination of the total X-ray mass and of the gas mass
fraction; this in turn allows to determine the level of the gas clumping
that can affect the shape of the entropy profiles reported by the
\textsl{Suzaku} observations. Our conclusion is that the role of the gas
clumping is very marginal and that the observed entropy flattening is due
to the rapid decrement of the temperature in the cluster outskirts caused
by non gravitational effects. Moreover, we show that the X-ray/SZ joint
analysis from \textsl{ROSAT} and \textsl{Planck} data, as performed in some
recent investigations, is inadequate to discriminate between a power law
increase and a flattening of the entropy.

\end{abstract}

\keywords{cosmic background radiation --- galaxies: clusters: individual
(Abell 1795, Abell 2029, Abell 2204, Abell 133) --- X-rays: galaxies:
clusters}

\section{Introduction}

The X-ray observations of \textsl{Suzaku} with a low and stable particle
background have started to shed some light on the essentially unexplored
outskirts of galaxy clusters, the sites of several interesting physical
processes and formation events (see Reiprich et al. 2013). The study of the
radial range $r_{500} - R$ that separates the virialized from the outer
infall region, is attracting an increasingly interest in cluster
cosmology\footnote{Here $r_{\Delta}$ is the radius within which the mean
density is $\Delta$ times the critical density, while $R$ is the viral radius
of the cluster. Frequently used values read $r_{500}\approx R/2$ and
$r_{200}\approx 3\, R/4$.}

Gas temperature profiles that strongly enter in the mass determinations are
found to decline beyond the central region ($\la 0.3\, r_{200}$) by a factor
of about three out to $r_{200}$ and slightly beyond (see Akamatsu et al.
2011; Reiprich et al. 2013). These profiles are rather similar for relaxed
and disturbed galaxy clusters. Of great interest are also the significant
variations observed in some azimuthal analysis. In the cool core clusters
Abell 1689 and Abell 1835 (Kawarada et al. 2010; Ichikawa et al. 2013) and in
the non-cool core Coma cluster (Simionescu et al. 2013) hot regions are
adjacent to filamentary structures, while the cold regions contact
low-density environments outside the clusters. These results suggest a more
efficient thermalization in the overdense infall regions.

A much discussed finding from the \textsl{Suzaku} observations of the cluster
outskirts regards the entropy profile (or rather the adiabat
$k=k_B\,T/n_e^{2/3}$) that shows a flattening above $\sim 0.5\,r_{200}$ (see
Walker et al. 2012c). In particular, the entropy flattening is more evident
when gas density profiles obtained by \textsl{Suzaku} are used. These
profiles exhibit systematic differences with respect to the \textsl{ROSAT}
density runs that appear to be steeper in the outskirts. The entropy shape
found in many clusters is in sharp contrast with the entropy profile $k
\propto r^{1.1}$ expected from pure gravitational infall (see Tozzi \& Norman
2001; Lapi et al. 2005; Voit 2005). The knowledge of the entropy profile is
fundamental to derive the intracluster plasma (ICP) structure and to obtain a
record of the thermal energy gains and radiative losses.

Walker et al. (2012c) derived the entropy profiles for a catalogue of relaxed
cool core clusters at redshift $\la 0.25$ studied with \textit{Suzaku},
adding to Abell 1835 and Abell 2204 investigated by \textsl{Chandra} in their
outskirts (Bonamente et al. 2013; Sanders et al. 2009). These authors
discussed the physical processes able to affect the entropy profiles. One
possibility to reconcile the observed entropy flattening with a power law
increase is constituted by the gas clumping expected in the ICP, as reported
by many hydrodynamical simulations. However, it is not still clear at which
level these inhomogeneities in the gas distribution are present, at which
distance from the cluster center they starts to be meaningful, and how they
behave radially. The clumping depends sensitively on the ICP physics, such as
the rate of cooling and star formation. Gas clumping implies an
overestimation of the gas density that would appear to drop less steeply,
with a consequent underestimation of the entropy and of the total mass.
Moreover, if the clumps are cool the temperature profile would appear to drop
more steeply, also concurring to an underestimation of the total mass. Each
of these effects leads to an overestimation of the gas mass fraction $f_{\rm
gas}$.

From simulated clusters, Mathiesen et al. (1999) reported a clumping factor
$C \equiv \langle\rho^2_{\rm gas}\rangle/\langle\rho_{\rm gas}\rangle^2 \sim
1.3-1.4$ inside $r_{500}$. Nagai \& Lau (2011) found that, at $r_{200}$, the
clumping $C$ takes on values $1.3-2$ depending on the presence or not of the
radiative cooling in the outskirts of their hydrodynamical simulations. Vazza
et al. (2013) report a radially increase of $C$ in all the simulated
clusters, in agreement with Nagai \& Lau (2011). The clumping is consistent
with $1$ in the innermost cluster regions and increases to values $\sim 3-5$
at the virial radius. Merging or post-merging clusters are on average
characterized by larger values of $C$ at all radii. The non-parametric method
used by Morandi \& Cui (2013) to measure inhomogeneities in the gas
distribution from X-ray observations reports for Abell 133, observed by
\textit{Chandra}, a radially increase of the gas clumping factor that reaches
$\sim 2-3$ at $r_{200}$, in good agreement with the predictions of
hydrodynamical simulations. When radiative cooling is included, Roncarelli et
al. (2013) find a very high level of clumpiness that ranges from $\sim 3$
close to the centre up to $\sim 10$ close to $r_{200}$. For the outer regions
$C$ attains values of $\sim 100$ at $2\,r_{200}$. Lower values ($C \sim 2-3$
at $r_{200}$) are obtained considering only the contribution of the emitting
gas. Zhuravleva et al. (2013) find that the typical value of the total
clumping factor in relaxed clusters varies from $\sim 1.2-1.6$ at $r_{500}$
up to $\sim 1.6-3.1$ at $1.5\,r_{500}$.

However, the values of $C$ reported in most of the simulations appear to be,
at least for some clusters, significantly lower than those necessary to
reconcile the derived gas mass fraction from X-ray observations with the
observed value from the CMB (e.g., Komatsu et al. 2011; \textsl{Planck}
Collaboration et al. 2013b) or to reconcile the observed entropy profiles
with the theoretically expected power law increase. From the \textsl{Suzaku}
analysis of the Perseus cluster, Simionescu et al. (2011) found that a
clumping factor $\sim 4-16$ over the radial range $0.7-1\,r_{200}$ is
required to make consistent the derived and measured gas mass fraction. Lower
values of $C$ for this cluster are obtained from the azimuthally resolved
X-ray spectroscopy of Urban et al. (2014). These values range from $\sim$ 1.2
to 2 or more at $r_{200}$ along different directions. In Abell 1835 a value
of $C\sim 7$ is necessary to make the entropy profile agree with a power law
increase in the outskirts (Walker et al. 2013; Fusco-Femiano \& Lapi 2013).
For PKS 0745-191 a value in the range $C\sim 2-9$ has been estimated at
$1.5\,r_{500}$ by Walker et al. (2012b).

Hoshino et al. (2010) and Akamatsu et al. (2011) have proposed that the
observed entropy flattening may be due to a possible difference between
electron and ion temperatures in accretion shocks. Thus, the flattening would
be the result of a lower electron temperature because the protons for their
higher mass thermalize first after the accretion shocks. However,
\textsl{Planck} observations of the pressure profiles in the cluster
outskirts (\textsl{Planck} Collaboration et al. 2013a) seem to exclude a
disequilibrium between electron and ions in these regions, for the lack of a
pressure drop. In addition, this conclusion appears to be reinforced by the
agreement between \textsl{Planck} gas pressure profiles with the simulation
outcomes (see also Wong \& Sarazin 2009).

Lapi et al. (2010) and Cavaliere et al. (2011) have proposed that the entropy
flattening results from a progressive saturation of the entropy production
during the late cluster growth, when the inflow across the virial boundary
peters out and the associated accretion shocks weaken. This occurs when the
accretion feeds on the tapering wings of a dark matter (DM) perturbation over
the background itself lowering under the accelerated cosmic expansion at low
$z$. The effect is enhanced in cluster sectors adjacent to low-density
regions of the surrounding environment, as it seems to be confirmed by the
significant azimuthal variations found in the aforementioned clusters.
Correspondingly, the weakening of the accretion shocks implies relatively
more kinetic energy to seep through the boundary, creating conditions
conducive to trigger turbulence in the ICP fluid (Cavaliere et al. 2011;
Cavaliere \& Lapi 2013).

Walker et al. (2012c) used the functional form reported in Lapi et al. (2012)
and Cavaliere et al. (2011) to fit the entropy profiles for a sample of
relaxed clusters at $z\la 0.25$ whose ICP has been studied out to $\sim
r_{200}$. This functional form fits the data well outside $0.3\, r_{200}$,
adding support to the suggestion that the flattening and downturn of the ICP
entropy can be the result of the weakened accretion. The authors have also
shown that the gas clumping calculated in the numerical simulations of Nagai
\& Lau (2011) is insufficient to reproduce the observed flattening and
turnover of the entropy. Moreover, they have shown that a temperature decline
much shallower than observed is necessary in the outskirts to reproduce a
power law increase of the entropy $k \propto r^{1.1}$, even using the gas
density profiles from \textsl{ROSAT}.

A different analysis to investigate the state of the ICP in the cluster
outskirts is based on the simultaneous use of X-ray and Sunyaev-Zel'dovich
(SZ) observations to model the density and temperature profiles. This
analysis exploits the possibility of easily obtaining the gas density profile
from the X-ray surface brightness in the soft band (0.5-2 keV) where the
temperature dependence is negligible in a hot cluster, and then use it to
obtain the ICP temperature from the SZ data. This avoids the difficulty of
measuring high-quality X-ray spectra (see Cavaliere et al. 2005). Of course,
any bias in the X-ray profile reflects immediately on the temperature
determination. For example, the eventual presence of gas clumping implies a
steeper decline in the temperature profile obtained by the X-ray/SZ joint
analysis.

This method has been used by Eckert et al. (2013a) to challenge the entropy
flattening reported in the \textsl{Suzaku} X-ray analysis of several clusters
(see Walker et al. 2012c). They used the average SZ electron pressure
profiles $P_e(r)$ from the \textsl{Planck} observations of $62$ clusters
(\textsl{Planck} Collaboration et al. 2013a) and the average \textsl{ROSAT}
gas density profiles (Eckert et al. 2012). The large field-of-view and the
low instrument background of \textsl{ROSAT} PSPC allow to reconstruct the
density profile out to the virial radius in $18$ clusters observed in common
with \textit{Planck}. These authors conclude that the entropy profiles $k(r)
= P_e(r)/n_e(r)^{5/3}$ agree with a power law increase expected from pure
gravitational infall.

It is well known that the traditional method to estimate the total X-ray
cluster mass $M(r)$ is based on the ICP density and temperature profiles,
that allow to solve the equation of hydrostatic equilibrium (HE) in spherical
symmetry. The X-ray masses result biased low by a systematic $\sim (10-20)\%$
even in relaxed clusters when compared with the strong and weak lensing
measurements (Arnaud et al. 2007; Mahdavi et al. 2008; Lau et al. 2009;
Battaglia et al. 2013). Moreover, the observed rapid decline of the
temperature leads to decreasing mass profiles in the outskirts of some
relaxed clusters (e.g., Kawaharada et al. 2010; Bonamente et al. 2013; Walker
et al. 2012b; Ichikawa et al. 2013) that may be explained in terms of an ICP
far from the HE, likely owing to the presence of a nonthermal gas pressure
support.

On the other hand, simulations agree in showing the presence of gas motions
caused by several processes (as inflow of material into the cluster from its
environment, mergers, and supersonic motions of galaxies through the ICP)
that may trigger the development of turbulence in the cluster outskirts
(Nagai et al. 2007; Shaw et al. 2010; Burns et al. 2010; Vazza et al. 2011;
Rasia et al. 2012). A nonthermal component may sustain the HE and resolve the
mass discrepancy discussed above, as shown by Fusco-Femiano \& Lapi (2013)
for Abell 1835. They exploited the possibility offered by the SuperModel
(Cavaliere et al. 2009) to include in its formalism a nonthermal pressure
component, and succeeded in reconstructing a total mass profile consistent
with the weak lensing measurements.

This paper is organized as follows. In the next \S~2 we briefly describe our
entropy-based SuperModel (SM). The temperature and the pressure profiles
include the contribution of a nonthermal, turbulent pressure component to
sustain the HE also in the cluster outskirts. Moreover, we report the
modified equation to compute the total X-ray mass $M(r)$ in presence of this
additional pressure component. In \S~3 we analyze the temperature profile of
four relaxed clusters (Abell 1795, Abell 2029, Abell 2204 and Abell 133)
exploiting observations by \textit{Suzaku}, \textsl{XMM-Newton} and
\textit{Chandra}. As gas density we use the \textsl{ROSAT} profiles reported
by Eckert et al. (2013a) for the former three clusters, and the
\textsl{Chandra} profile for Abell 133 (Morandi \& Cui 2013). We also perform
the SM analysis of the \textsl{Planck} pressure profiles, obtained by
observations of the SZ effect (\textsl{Planck} Collaboration et al. 2013a).
We discuss the results and draw our conclusions in \S~4.

Throughout the paper we adopt the standard flat cosmology with parameters in
round numbers: $H_0 = 70$ km s$^{-1}$ Mpc$^{-1}$, $\Omega_{\Lambda} = 0.7$,
$\Omega_M = 0.3$ (Komatsu et al. 2011; Hinshaw et al. 2013; \textsl{Planck}
Collaboration 2013b). Then 1 arcmin corresponds to $71.75$ kpc for Abell
1795, to $88.60$ kpc for Abell 2029, to $158.97$ kpc for Abell 2204 and to
$65.92$ kpc for Abell 133.

\section{SuperModel with Turbulence}

In our SM, the profiles of ICP density $n$ and temperature $T$ are obtained
via the HE equation by specifying the entropy distribution $k = k_B\,
T/n^{2/3}$. The central entropy levels are set by the balance between
production processes like AGN outbursts and deep mergers, versus the erosion
process by radiative cooling. In the outskirts, close to the virial radius
$R$ of the cluster, entropy is produced by supersonic inflows of gas from the
environment into the DM gravitational potential well. Thence the ICP entropy
adiabatically stratifies, yielding a spherically averaged profile with shape
$k(r) = k_c + (k_R - k_c)(r/R)^a$, see Voit (2005). The central floor
$k_c\approx 10-100$ keV cm$^2$ goes into a power law increase with slope $a
\approx 1.1$ (Tozzi \& Norman 2001; Lapi et al. 2005) out to the boundary
values $k_R \sim$ some $10^3$ keV cm$^2$.

Steep temperature and flat entropy profiles have been observed by
\textsl{Suzaku} toward the virial radius in some relaxed CC clusters, and in
the undisturbed directions of NCC clusters (like Coma). These findings can be
interpreted in terms of a reduced entropy production relative to a pure
gravitational inflow. In fact, the latter occurs when the accretion rates
peters out and the accretion shocks weaken due to the slowdown at low $z$ of
the cosmological structure growth in an accelerating Universe (particularly
evident in relaxed, CC clusters). The effect is more pronounced in cluster
sectors (both of CC and NCC clusters) adjacent to low-density regions of the
surrounding environment, implying azimuthal variations of the X-ray
observables.

This behavior is modeled in the SM through an entropy run that starts as a
simple powerlaw with slope $a$, but for radii $r > r_b$ deviates downward
(Lapi et al. 2010). For the sake of simplicity, the entropy slope is taken to
decline linearly with a gradient $a^{\prime} \equiv (a-a_R)/(R/r_b - 1)$,
where $r_b$ and $a^{\prime}$ are free parameters to be determined from the
fitting of the X-ray observables.

The weakening of the accretion shocks is also expected to let relatively more
bulk energy to seep through the cluster, and drive turbulence into the
outskirts (Cavaliere et al. 2011). Turbulent motions originate at the virial
boundary with a coherence lengths $L \sim R$/2 related to the pressure scale
height or to shock segmentation (see Iapichino \& Niemeyer 2008; Valdarnini
2011; Vazza et al. 2010), and then they fragment downstream into a dispersive
cascade to sizes $l$. Numerical simulations show that small values of the
turbulent energy apply in the cores of relaxed clusters, but the ratio
$E_{\rm turb}/E_{\rm thermal}$ of the turbulent to thermal energy increases
into the outskirts (e.g., Vazza et al. 2011).

In the presence of turbulence, HE is sustained not only by thermal pressure,
but also by an additional nonthermal contribution due to turbulent motions;
the latter features a radial shape decaying on the scale $l$ from the
boundary, outer value $\delta_R$. The total pressure can be written as
$p_{\rm tot}(r) = p_{\rm th}(r) + p_{\rm nth}(r) = p_{\rm th}(r)[1 +
\delta(r)]$ in terms of the quantity $\delta(r) \equiv p_{\rm nth}/p_{\rm
th}$. The HE equation yields the temperature profile as
\begin{equation}
\frac{T(r)}{T_R} = \left[\frac{k(r)}{k_R}\right]^{3/5}\, \left[\frac{1 +
\delta_R}{1 + \delta(r)}\right]^{2/5}\,\left\{1 + \frac{2}{5}\frac{b_R}{1 +
\delta_R}\int_r^R {\frac{{\rm d}x}{x} \frac{v^2_c(x)}{v^2_R}\,
\left[\frac{k_R}{k(x)}\right]^{3/5}\, \left[\frac{1 + \delta_R}{1 +
\delta(x)}\right]^{3/5}}\right\}
\end{equation}
and the pressure profile as
\begin{equation}
\frac{P(r)}{P_R} = \left[\frac{1 + \delta_R}{1 + \delta(r)}\right]\left\{1 +
\frac{2}{5} \frac{b_R}{1 + \delta_R}\,\int_r^R {\frac{{\rm d}x}{x}
\frac{v^2_c(x)}{v^2_R} \left[\frac{k_R}{k(x)}\right]^{3/5}\, \left[\frac{1 +
\delta_R}{1 + \delta(x)}\right]^{3/5}}\right\}^{5/2}
\end{equation}
where $v_c$ is the DM circular velocity ($v_R$ is the value at the virial
radius $R$), and $b_R$ is the ratio at $R$ of $v^2_c$ to the sound speed
squared (Cavaliere et al. 2009).

In our SM analysis we use the functional shape
\begin{equation}
\delta(r) = \delta_R\, e^{-(R-r)^2/l^2}
\end{equation}
which decays on the scale $l$ inward of a round maximum. This profile of
$\delta(r)$ concur with the indication of numerical simulations (Lau et al.
2009; Vazza et al. 2011). We remark that Morandi et al. (2012; see also Shaw
et al. 2010) adopted instead a power law for the fraction $p_{\rm nth}/p_{\rm
tot}$ in their 3-D structure reconstruction of Abell 1835.

The traditional equation to estimate the total X-ray mass $M(r)$ within $r$
is modified as follows to take into account the additional nonthermal
pressure component (Fusco-Femiano \& Lapi 2013)
$$M(r) = - \frac{k_B [T(r)(1 +\delta(r)] r^2 }{\mu m_p G}\left\{
\frac{1}{n_e(r)}\frac{d n_e(r)}{d r} + \frac{1}{T(r)[(1+\delta(r)]}\frac{d T(r)[1 +
\delta(r)]}{d r}\right\} $$
\begin{equation}
= - \frac{k_B [T(r)(1 +\delta(r)] r^2}{\mu m_p G}\left[\frac{1}{n_e(r)}
\frac{d n_e(r)}{d r} + \frac{1}{T(r)}\frac{d T(r)}{d r} + \frac{\delta(r)}{1
+ \delta(r)} \frac{2}{l^2}(R - r)\right]~.
\end{equation}
The hot gas mass writes
$$M_{\rm gas} = 4\pi \mu_e m_p\int{\rm d}r~{n_e(r) r^2}$$
where $\mu_e \sim 1.16$ is the mean molecular weight of the electrons.

\section{SuperModel Analysis for Abell 1795, Abell 2029, Abell 2204 and Abell 133}

Here we perform the SM analysis of the temperature profile of four relaxed
clusters considering two entropy profiles, namely a power law increase and an
entropy run that starts with an initial slope $a$, and then deviates downward
when $r > r_b$. For $n_e(r)$ we use the \textsl{ROSAT} gas density profiles
(Eckert et al. 2013a) that are found to be steeper in the cluster outskirts
than the \textsl{Chandra} and \textsl{Suzaku} profiles, implying lower gas
mass fractions and larger entropy. The virial radius $R$ is assumed to be
$2r_{500}$ where the radius $r_{500}$ is reported in Table 1 of
\textsl{Planck} Collaboration et al. (2011), and has been calculated
iteratively as described in Kravtsov et al. (2006). For Abell 133 we consider
the temperature and density profiles derived by the \textsl{Chandra} analysis
of Morandi \& Cui (2013) and the density profile that they obtain when the
inhomogeneities of the gas distribution are taken into account. We also
analyze the pressure \textsl{Planck} profiles (\textsl{Planck} Collaboration
et al. 2013a) showing the inadequacy in the use of the pressure to determine
the entropy profile via the relation $k(r) = P(r)/n_e(r)^{5/3}$.

\subsection{Abell 1795}

Abell 1795 appears quite regular and relaxed with a cool core, although
\textsl{Chandra} has evidenced a plume associated to its brightest cluster
galaxy (Fabian et al. 2001). Bautz et al. (2009) have estimated from the
observed temperature (spatially averaged $\sim$ 5.3 keV) $r_{500} = 1.3~
h^{-1}_{70}$ Mpc and $r_{200} = 1.9~ h^{-1}_{70}$ Mpc. At the redshift of the
cluster ($z$ = 0.063) the virial radius is larger than $r_{200}$. They expect
that $R \approx 1.35\,r_{200} \approx$ 2.56 Mpc in agreement with the value
$R = 2r_{500}$ = 2.51 Mpc ($\sim 35^{\prime}$) that we assume.

Our SM fit to the temperature profile considers the runs in the north and
south cluster sectors observed by \textsl{Suzaku} (Bautz et al. 2009) along
with results from \textsl{XMM-Newton} (Snowden et al. 2008). We do not
consider the \textsl{Chandra} data (Wikhlinin et al. 2006) that are higher
than the \textsl{Suzaku} and \textsl{XMM-Newton} temperatures outside of the
cool core (see Bautz et al. 2009 for more details). The analysis starts
assuming only thermal pressure for the HE ($\delta(r) \equiv p_{\rm
nth}/p_{\rm th} = 0$) and an entropy profile that flattens at $r > r_b$ (blue
line of Fig.~1). The dashed green line is instead for an entropy profile that
follows a power law. It is evident that this steep temperature profile is
consistent only with an entropy run that deviates downward at $r > r_b$. The
gas density profile is obtained by the SM fit to the \textsl{ROSAT} data
(Eckert et al. 2013a) with $n_e \sim 1.21 n_H$ (Fig.~1).

The rapid decline of the temperature profile leads to a decreasing mass
profile in the cluster outskirts and a consequent gas mass fraction well
above the cosmic value at the virial radius (blue lines of Fig.~2). The
decreasing mass profile suggests the presence of a nonthermal, turbulent
pressure component that adds to the thermal in sustaining the cluster HE. To
reconstruct the total X-ray mass of Abell 1835 (Fusco-Femiano \& Lapi 2013)
the quantities $\delta_R$ and $l$ (see Eq.~3) have been determined from the
SM fits to the temperature and brightness profiles imposing that the baryon
mass fraction equals the cosmic value at the virial radius, and that the mass
profile is smooth in the outskirts. The result is a X-ray virial mass
consistent with the weak lensing mass measured by Hoekstra et al. (2012). The
constraint $f_{\rm gas} = \Omega_b/\Omega_M - f_{stars}$ at $R$ is supported
by X-ray and SZ observations. \textsl{Suzaku} reports for Abell 1835
(Ichikawa et al. 2013) a gas mass fraction, defined by the total lensing
mass, agreeing at the virial radius with the cosmic baryon fraction.
Moreover, the combined analysis of Eckert et al. (2013b) shows that, at
$r_{200}$, $f_{\rm gas}$ converges for relaxed clusters to the expected
value. Also the \textsl{Planck} constrains are compatible with the cosmic
value at large radii (\textsl{Planck} Collaboration et al. 2013a).

For Abell 1795 the total X-ray and gas masses that determine $f_{\rm gas}$
are obtained by the SM fit to the temperature data for given values of
$\delta_R$ and $l$ (red line of Fig.~1) and by the \textsl{ROSAT} gas
density. The above constraints are satisfied for $\delta_R$ = 1.3 and $l$ =
0.5 that yield the deprojected temperature profile shown by the red line of
Fig.~3; blue line is for $\delta$ = 0.
The higher contribution to
$f_{\rm gas}$ by inhomogeneities in the ICP distribution may be derived
imposing a flat profile to the X-ray mass in the cluster outskirts obtained
with $\delta_R$ = 1.1 (dashed black line in Fig.~2, left panel). The consequent
slight discrepancy between $f_{\rm gas}$ and the cosmic value (dashed black
line of Fig.~2, right panel) can be attributed to a clumping gas factor $C
\sim$ 1.3 at the virial boundary. The knowledge of the virial lensing mass
($M_R^{lens}$) would allow to quantify the value of $C$. If $M_R^{lens}$
agrees with the value of the total X-ray mass at $R$ given by the red profile
the gas mass fraction equals the cosmic value at the virial radius (red line)
making null the contribution to $f_{\rm gas}$ by gas clumping. It can reach
the value of $\sim$ 1.3 if $M_R^{lens}$ is consistent with the flat X-ray
mass profile. Thus $C\leq$ 1.3. This upper limit for $C$ can raise the red
entropy profile (see Fig.~3) at most of a factor $C^{1/3} \sim 1.1$ at $R$
clearly insufficient to explain the entropy flattening (Fig.~3, right
panel). In the case of Abell 1835 the X-ray mass profile that satisfies the
condition $f_{\rm gas} = \Omega_b/\Omega_M - f_{stars}$ at the virial radius
is consistent with the measured $M_R^{lens}$ giving $C \simeq$ 0. Following
our SM analysis of Abell 1795 we predict a virial mass of $(8 - 9) \times
10^{14}$ $M_{\bigodot}$ as shown by Fig.~2 (left panel).

A different approach to investigate the state of the ICP in cluster outskirts
is based on the simultaneous use of X-ray and SZ observations to model the
density and temperature profiles (Cavaliere et al. 2005). In Fig.~4 we have
performed a SM fit to the \textsl{Planck} pressure profile of Abell 1795
(\textsl{Planck} Collaboration 2013a) considering a power law increase of the
entropy (dashed line) and an entropy run that deviates downward at $r > r_b$
(continuous line), as used above for the temperature profile. It results a
moderate gap between the two curves well inside the pressure error bars. The
derived entropy profiles (see Fig.~4) are characterized by large
uncertainties making this approach unsuitable to discriminate between the two
profiles. This result reflects the weak dependence of the pressure on the
entropy (Eq.~2) at variance with the much stronger dependence of the
temperature on $k$ (Eq.~1). The prevalence of the relation $k = T/n_e^{2/3}$
with respect to $k = P/n_e^{5/3}$ in determining the entropy profile is also
evident when we compute the gradients of the pressure and of the temperature
at the virial radius for the two different entropy profiles:
\begin{equation}
g_P \equiv (\frac{dlnP }{dln r})_R = -\frac{b_R}{1+\delta_R}
\end{equation}
\begin{equation}
g_T \equiv (\frac{dlnT }{dln r})_R = \frac{3}{5}a_R-\frac{2b_R}{5(1+\delta_R)}
\end{equation}
where $b_R = (45-19a_R)/9$ (Cavaliere et al. 2009) and $a_R = a-(R/r_b
-1)a^{\prime}$ (Lapi et al. 2010). Following the above relation for $a_R$,
the two entropy profiles adopted in our SM analysis are characterized by
$r_b$ = R (power law) and $r_b < R$ (entropy flattening). For simplicity we
assume $\delta_R$ = 0, $a$ = 1.1 and $a^{\prime}$ = 0.5. It results that for
$r_b = R$ and $r_b = 0.3R$ the relative variation for the pressure is
$\Delta_P = (g_P^{r_b = R} - g_P^{r_b = 0.3R})/g_P^{rb = R}\sim$ -0.92 while
for the temperature $\Delta_T \sim$ -4.10.

\subsection{Abell 2029}

The intracluster medium of Abell 2029 ($z = 0.0767$) has been investigated by
\textsl{Suzaku} at radii near the virial boundary and with a good azimuthal
coverage (Walker et al. 2012a). It appears a relaxed cluster reported also by
several previous X-ray observations with \textsl{ASCA} and \textsl{ROSAT}
(Sarazin et al. 1998), \textsl{Beppo-SAX} (Molendi \& De Grandi 1999),
\textsl{Chandra} (Vikhlinin et al. 2006) and \textsl{XMM-Newton} (Bourdin \&
Mazzotta 2008; Snowden et al. 2008). It belongs to a small supercluster with
other three members (A2033, A2028 and A2066) giving the opportunity to study
the influence of a such environment on the cluster outskirts.

The \textsl{Suzaku} observations report that the temperature and the entropy
are lower in the SE than in other directions which are consistent with each
other. The second asymmetry is an excess in the north of the projected
density above the azimuthal average likely due to a filamentary structure
connecting Abell 2029 with the closest cluster Abell 2033. The
\textsl{Suzaku} temperature points of Fig.~5 are azimuthally averaged
excluding the SE and the north. In this figure we report our SM analysis that
starts assuming that the pressure is only thermal ($\delta$ = 0). A better
fit is obtained for an entropy profile that deviates from a simple power law
increase. Our total mass profile (Fig.~5) slightly decreases going toward the
virial radius ($R = 2r_{500}\sim 31.4^{\prime}$, see \textsl{Planck}
Collaboration 2011) causing an increase of the gas mass fraction above the
cosmic value (Fig.~6, continuous blue line). The SM X-ray mass profile is
consistent with the mass value at $r_{500}$ reported by the
\textsl{XMM-Newton} analysis of Gonzalez et al. (2013); instead their gas
mass value derived by the brightness surface is slightly higher than the
\textsl{ROSAT} profile (see Fig.~6).

A higher value of $f_{\rm gas}$ is obtained using the shallower and higher
gas density profile of \textsl{Suzaku} (Walker et al. 2012a) reported in Fig.~7.
This value is due to the combined effect of a lower total X-ray mass for
the lower derivative of the density (see dashed blue line in Fig.~5) and to
an increase of $M_{\rm gas}$. The total X-ray mass profile obtained with the
\textsl{Suzaku} gas density profile and with the SM fit to the
\textsl{Suzaku} temperature profile is consistent with the value of
$8.0^{+1.5}_{-1.5}\times 10^{14}M_{\bigodot}$ derived at $r_{200} =
22.0^{+1.3}_{-1.4}$ arcmin by Walker et al. (2012a) (see Fig.~5). Moreover,
the higher density profile explains the lower entropy values derived by the
\textsl{Suzaku} analysis of Abell 2029 with respect to the SM entropy profile
obtained with the \textsl{ROSAT} gas density profile (Fig.~6).

The decreasing mass profile beyond $r_{200}$ suggests the presence of
turbulence to sustain the HE. With $\delta_R$ = 0.5 and $l$ = 0.5 we obtain
an increasing mass profile that allows to match $f_{\rm gas}$ to the cosmic
value (see red line in Fig.~6). The use of the \textsl{Suzaku} gas density
profile requires a higher level of turbulence for the higher gas mass
fraction value. To evaluate the possible contribution of the gas clumping we
consider a flat profile for the X-ray mass (black dashed line in Fig.~6;
$\delta_R$ = 0.3 and $l$ = 0.5). This mass profile implies $C \leqslant$ 1.3
and therefore a slight increase of a factor $C^{1/3} \leqslant$ 1.1 of the
entropy at $R$ insufficient, as for Abell 1795, to explain the entropy
flattening. From our SM analysis we predict a virial mass of $(1.2 - 1.3)
\times 10^{15} M_{\bigodot}$ (see Fig.~5). Also for Abell 2029 the fits to
the \textsl{Planck} pressure profile with and without entropy flattening are
within the uncertainties of the data (Fig.~7).

\subsection{Abell 2204}

The regular cluster Abell 2204 has been observed by \textsl{Suzaku} out to
$\sim$ 1800 kpc (Reiprich et al. 2009). This distance is close to an estimate
of $r_{200}\sim$ 1840 ($\sim 11.7^{\prime}$) obtained by extrapolating the
mass profile derived by the \textsl{XMM-Newton} analysis of Zhang et al.
(2008). Here we assume $R = 2r_{500} \sim$ 16.92 arcmin. Our SM analysis
starts with the fit to the \textsl{Suzaku} and \textsl{XMM-Newton}
temperature data with $\delta$ = 0 (see Fig.~8). We disgregard the
\textsl{Chandra} data (Reiprich \& Bohringer 2002) that are higher than the
values of the other two X-ray observatories at $r < 3^{\prime}$. A better fit
is obtained with a deviation of the entropy from a simple power law increase.
With the gas density obtained by the fit to the \textsl{ROSAT} observations
(Eckert et al. 2013a) we trace the X-ray mass profile of Fig.~8 that slightly
increases in the cluster outskirts and is found to be consistent with the
\textsl{XMM-Newton} profile within $r_{500}$ (Zhang et al. 2008). This figure
reports also the best fit with a Navarro, Frenk \& White (NFW) or a singular
isothermal sphere (SIS) models to the weak lensing data (Clowe \& Schneider
2002). The mass profile given by the NFW model requires a nonthermal pressure
component of (10 - 15)\% of the total pressure at the virial radius to
reconcile the X-ray with the weak lensing mass, while larger values are
necessary using the SIS model. The former gives an increasing gas mass
fraction profile above the cosmic value at $R$, and the opposite holds for
the second model (see Fig.~9). The NFW mass profile implies a clumping factor
of $\sim$ 1.8 to reconcile $f_{\rm gas}$ with the observed value. This value
of $C$ represents an upper limit because it progressively decreases for
larger values of the weak lensing mass at the virial boundary. However, this
upper limit for $C$ is unable to explain the entropy flattening of Fig.~9
that requires $C\sim$ 8.2, confirming that the observed entropy flattening is
mainly due to the low temperatures in the outskirts rather than to gas
clumping. Also for Abell 2204 the fits to the \textsl{Planck} pressure
profile with the two entropy profiles are within the error bars (Fig.~10).

\subsection{Abell 133}

Abell 133 is a cool core galaxy cluster ($z$ = 0.057) deeply investigated by
\textsl{Chandra} (Vikhlinin \textit{et al.}, in prep) with several pointings
at distances near $r_{200}$. It is considered an optimal cluster by Morandi
\& Cui (2013) to apply their method and derive gas inhomogeneities at large
radii exploiting the excellent angular resolution of \textsl{Chandra} to
distinguish emission by clumps or by diffuse gas (Morandi et al. 2013).

In their paper Morandi \& Cui (2013) report the projected and deprojected
temperature profiles, the electron density $n_e$ obtained by deprojecting the
surface brightness profile and the derived entropy profile. With their
approach, based on the imprints left by the inhomogeneities of the gas on the
surface brightness distribution, they are able to trace the profile of the
gas clumping factor $C$, finding it in good agreement with the predictions of
hydrodynamical simulations (e.g., Nagai \& Lau 2011); they thus derive the
density and entropy profiles corrected for the effect of the gas
inhomogeneities. From their analysis $r_{200}$ = 1596$\pm$29 kpc applies, a
value that we assume as the virial radius $R$ in our SM analysis in order to
be very close to the boundary radius of 1500 kpc adopted by the authors in
their \textsl{Chandra} analysis. In this way we obtain a deprojected
temperature profile (see Fig.~11) consistent with their profile (Fig.~3 of
Morandi \& Cui 2013). For the SM fit to the temperature profile we consider
the two entropy profiles that we have adopted for the previous three
clusters. An evident better fit is obtained for the entropy profile that
deviates at $r > r_b$ from a power law increase. The X-ray cluster mass,
reported in Fig.~12, shows a slight decline near $r_{200}$ giving a gas mass
fraction value slightly above the cosmic value (see Fig.~12, blue lines).
This cannot be explained by the gas clumping factor estimated by Morandi \&
Cui (2013) because their modified gas density profile gives a gas mass
fraction well below the observed value (black line of Fig.~12). A similar
conclusion is obtained assuming $R = 4/3r_{200}$ as for the previous
clusters. Instead a modest level of turbulence ($p_{\rm nth} \sim 5\% p_{\rm
tot}$ at $R$) is sufficient to make $f_{\rm gas}$ consistent with the cosmic
value (see Fig.~12) and to obtain a non decreasing mass profile (red lines).
In Fig.~13 we plot the SM entropy profile quite consistent with that derived
by Morandi \& Cui (2013), but not with a power law increase even considering
the gas density corrected for the clumping effect.

\section{Discussion and Conclusions}

Several physical processes believed to occur in the cluster outskirts can be
constrained by the study of the ICP thermodynamic properties. An important
contribution is given by the \textsl{Suzaku} and \textsl{Planck} observations
in these peripheral cluster regions.

As reported in the Introduction, one of the most interesting findings of the
\textsl{Suzaku} observations is the deviation of the entropy profile at $r >
r_b$ (see Walker et al. 2012c) from the expected power law increase ($k \sim
r^{1.1}$) for pure gravitational infall (Voit et al. 2005). This
entropy flattening raises the following question: is it due to the observed
steep decline of the temperature at $r \ga 0.3\,r_{200}$ caused by
non-gravitational effects (e.g., Lapi et al. 2010), or to the presence of gas
clumping that implies an overestimation of the gas density in the cluster
outskirts with a consequent underestimation of the entropy (e.g., Nagai \&
Lau 2011; Vazza et al. 2012) ? On the other hand, the presence of any entropy
flattening has been challenged by Eckert et al. (2013a) with the simultaneous
use of X-ray and SZ observations ($k = P/n_e^{5/3}$) that allows to avoid the
difficulty in obtaining high quality X-ray spectra in the cluster outskirts.
As discussed in the Introduction, a negative aspect of this approach is that
any bias in the X-ray surface brightness reflects on the temperature
determination. Steeper declines may be obtained in presence of gas clumping.

Eckert et al. (2013a) found slightly flatter entropy profiles for NCC
clusters, while for CC clusters the entropy profiles are in excess compared
to the predicted power law increase out to $r_{500}$ and then converge at
larger radii. They contest to Walker et al. (2012c), that confirmed their
results using the same approach based on the X-ray/SZ joint analysis, to have
mixed CC and NCC clusters without considering that NCC systems are in
majority both in the \textsl{Planck} and \textsl{ROSAT} samples. Moreover,
they stress that the normalization for the entropy profiles with
$k(0.3\,r_{200})$ is arbitrary. But Walker et al. (2013) finds that the
entropy flattening can be confirmed also normalizing the entropy at
$r_{500}$. Within this radius the entropy is in excess of the expected power
law increase for most of the clusters, while outside the entropies are
systematically below the baseline prediction using only the gravitational
collapse. They suggest that gas clumping may be a possible explanation. The
same approach has been used by Pratt et al. (2010) for the REXCESS sample of
clusters observed with \textsl{XMM-Newton} that show a similar behaviour of
the entropy inside $r_{500}$.

One possibility to ascertain the presence of a flattening in the entropy
profile and, if this the case, to individuate the main process that
determines it is given by the SuperModel, the only tool based on the run of
the ICP specific entropy $k = k_B T/n^{2/3}$. Moreover, the SM temperature
profile (see Eq.~1) obtained when a nonthermal pressure component is inserted
in the HE equation allows to derive accurate X-ray cluster masses (see
Eq.~4).

The SM analysis of the X-ray and \textsl{Planck} observations of our small
sample of relaxed clusters is based on two entropy profiles, namely a power
law increase ($k \sim r^a$) and an entropy profile that starts with a power
law increase with slope $a$ and then deviates downward at radii greater than
$r_b$. The use of these two entropy profiles allows to show the inadequacy of
the method based on the joint X-ray/SZ analysis ($k = P/n_e^{5/3}$) in
determining the presence of an entropy flattening. The fits to the
\textsl{Planck} pressure profiles are within the error bars for both the
entropy profiles that we consider in our SM analysis highlighting the
difficulty to individuate the correct entropy shape. Such difficulty arises
from the very weak dependence of the pressure on $k$ as shown by Eq.~2. The
pressure and temperature gradients at the virial radius (see Eqs.~ 5 and 6)
computed for the two entropy profiles clearly indicate the prevalence of the
relation $k = T/n_e^{2/3}$ with respect to $k = P/n_e^{5/3}$ in determining
the entropy profile. Also Fig.~4 of Walker et al. (2012c) reflects the weak
dependence of $P$ on $k$. The baseline entropy $k \sim r^{1.1}$, normalized
at $0.3\,r_{200}$, is within the large scatter out to $\sim 0.9\,r_{200}$ of
the entropy profile obtained by combining the \textsl{Planck} pressure
profile derived from a sample of 62 clusters (\textsl{Planck} Collaboration
et al. 2013a) and the density profile derived with \textsl{ROSAT} PSPC from a
sample of 31 clusters (Eckert et al. 2012). The effect is similar or even
more evident for a normalization at $r_{500} \sim 0.66\,r_{200}$ (Walker et
al. 2013).

We stress that our SM analysis does not require to define a normalization for
the entropy profiles. These are obtained by combining the gas density of
\textsl{ROSAT} with the fits to the temperature profile obtained either with
a power law increase of the entropy or with a profile of $k$ that deviates at
$r > r_b$. The latter profile gives better fits to $T(r)$ for all the
clusters of our sample. For the relaxed clusters here investigated we do not
find the presence of an entropy excess within $r_{500}$ compared to a power
law profile of $k$. This excess is reported by Walker et al. (2013) with a
normalization at $k(r_{500})$ not present instead with a normalization at
$k(0.3\,r_{200})$ (Walker et al. 2012c). The deviation of the entropy from a
power law increase ($k \sim r^a$) is at $r \ga (0.3-0.4)R \approx
(0.4-0.5)\,r_{200}$. The value of $a$ is consistent with $\sim$ 1.1 for all
the clusters of our sample. We find low central levels $k_c \la$ 15 keV
cm$^2$ typical of CCs. High central floors ($k_c \approx 3\times 10^2$ keV
cm$^2$) are found in most of the NCC clusters (Fusco-Femiano et al. 2009;
Cavagnolo et al. 2009; Pratt et al. 2010).

Our analysis allows to derive upper limits to the gas clumping factor $C$;
this is because the SuperModel can easily include in its formalism the
contribution of turbulent pressure components, fundamental to sustain
equilibrium in cluster outskirts. In this way accurate X-ray masses and
$f_{\rm gas}$ values can be obtained. As shown in Sect. 3, the knowledge of
the weak lensing mass at the virial radius would allow to fix the value of
$C$. However, we have shown that these upper limits are absolutely
insufficient to justify the observed entropy flattening for our sample of
galaxy clusters. We believe that this modest presence of gas clumping is
consistent with the use of the gas density profiles observed by
\textit{ROSAT}. These profiles, steeper than those reported by
\textsl{Chandra} and \textit{Suzaku}, could be not affected by a significant
presence of gas clumping, predicted by hydrodynamical simulations, for the
poor spatial resolution of \textsl{ROSAT} that smears out any clumpy
emission, leading to smooth gas density profiles.

The conclusion is that the entropy flattening is due to the rapid decline of
the temperature observed by \textsl{Suzaku} in several cluster outskirts.
Lapi et al. (2010) and Cavaliere et al. (2011) have suggested that in relaxed
clusters the slowdown of the entropy production is originated by the
weakening of accretion shocks. The inflows through the boundary dwindle away
as they draw on the tapering wings of the initial perturbation over the
background provided by the accelerating universe (see also Cavaliere \& Lapi
2013). In such clusters weaker boundary shocks prevail and let relatively
more bulk inflow energy seeps through, ready to drive more turbulence. The
decreasing thermalization is more pronounced in cluster sectors adjacent to
low density regions of the surrounding environment. This agrees with the
azimuthal variations reported in some CC clusters by \textsl{Suzaku}
(Ichikawa et al. 2013) and in the NCC Coma cluster (Simionescu et al. 2013).
The latter cluster shows in such sectors entropy profiles consistent with
those found in CC clusters. Following the above interpretation this
consistency indicates that the rate and strength of the accretion shocks
along undisturbed directions are similar in the NCC Coma cluster and in the
more evolved CC clusters.

Other explanations have been suggested for the flattening of the entropy
profiles. One possibility involves gas clumping at large radii (Simionescu et
al. 2011). This process is not supported by the SM analysis of the relaxed
clusters here investigated and of Abell 1835 (Fusco-Femiano \& Lapi 2013)
that gives upper limits of $C$ too low to report the derived entropy profiles
to the predicted power law increase. A similar conclusion is for four
clusters examined by Walker et al. (2013) where the entropy flattening is
attributed to the sharp decrement of the temperature in their outskirts. One
of these clusters is Abell 1835. Instead for three clusters and one of these
is Abell 2029 the overdensity appears the cause of the entropy flattening. We
stress that the overdensity reported in their Fig.~13 is mostly due to the
use of the \textsl{Suzaku} density profile instead of the steeper
\textsl{ROSAT} profile. However, we have shown that also using the
\textsl{ROSAT} observations the low upper limit of the gas clumping factor
derived for this cluster is insufficient to reconcile the observed entropy
flattening with a power law increase (see Fig.~6). An alternative explanation
for the entropy flattening is based on the electron-ion non-equilibrium in
the cluster outskirts proposed by Hoshino et al. (2010) and Akamatsu et al.
(2011). As observed by Simionescu et al. (2013) in this case the similar
shapes of the entropy profiles in the merging Coma cluster and in CC clusters
would require a similar age and strength of the last shock that the gas has
experienced in the outskirts. However, this disequilibrium seems excluded for
the lack of a pressure drop in the \textsl{Planck} observations
(\textsl{Planck} Collaboration et al. 2013a).

In conclusion, the discrepancy between the results of Walker et al. (2012c;
2013) and Eckert et al. (2013a) regarding the shape of the entropy profile is
due to the weak dependence of the pressure used in the X-ray/SZ joint
analysis on the entropy $k$. Exploiting the ability of the SuperModel to
include in its formalism a nonthermal pressure component sustaining the HE it
is possible to obtain accurate X-ray masses and thus reliable gas mass
fractions. The derived contribution of the gas clumping to $f_{\rm gas}$
results in our sample not decisive to give a power law increase of the
entropy. This implies that the entropy flattening derived by \textsl{Suzaku}
observations in several CC clusters and in the dynamically active Coma
cluster is due to the rapid decline of the temperature in the cluster
outskirts. The azimuthal variations of the ICP thermodynamic properties found
in these clusters imply a decreasing thermalization, more pronounced in
cluster sectors adjacent to low-density regions of the surrounding
environment or in the undisturbed directions of the cluster outskirts. The
weakening of the accretion shocks that leads to a slowdown of the entropy
production may be a plausible explanation for the scenario outlined by the
\textsl{Suzaku} observations.

\begin{acknowledgements}
We thank our referee for constructive comments. This work has been supported
in part by the MIUR PRIN 2010/2011 ‘The dark Universe and the cosmic
evolution of baryons: from current surveys to Euclid’, by the INAF PRIN
2012/2013 ‘Looking into the dust-obscured phase of galaxy formation through
cosmic zoom lenses in the Herschel Astrophysical Terahertz Large Area
Survey’. A.L. thanks SISSA for warm hospitality.
\end{acknowledgements}

\clearpage
\begin{figure*}
\begin{center}
\epsscale{1.15}\plottwo{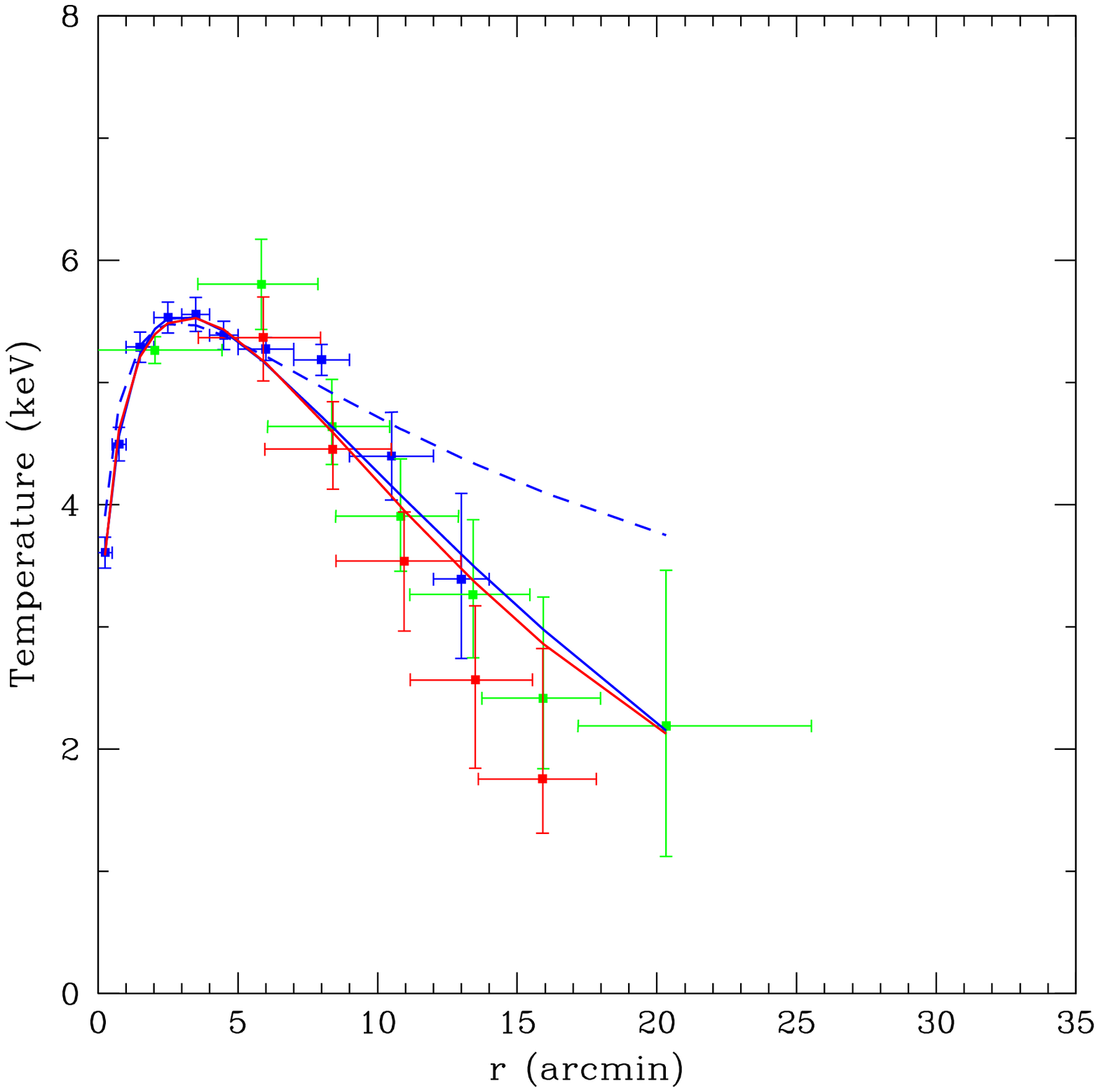}{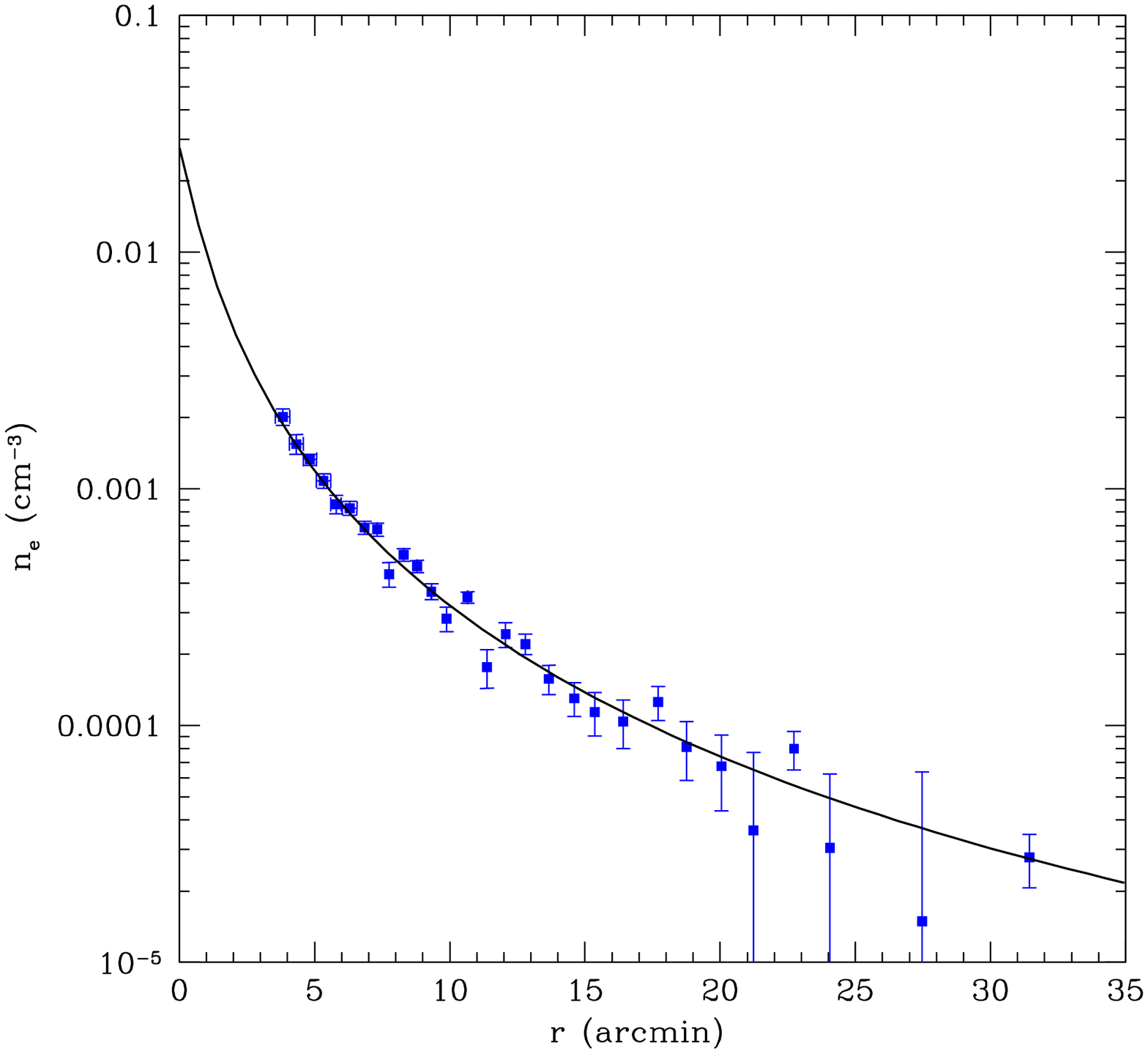}\caption{Abell
1795 - Left panel: Projected temperature profiles. Green and red points from
\textsl{Suzaku} (Bautz et al. 2009). They refer to the North and Sud sectors,
respectively; blue points from \textsl{XMM-Newton} (Snowden et al. 2008).
Blue and red lines are the SM fits without ($\delta_R$ = 0) and with
turbulence ($\delta_R$ = 1.3, $l$ = 0.5), respectively. Both the lines are
obtained with an entropy profile that deviates from a power law at $r > r_b$;
dashed blue line is the SM fit given by a power law increase of the entropy
profile. Right panel: Black line is the SM fit to the electron density
points from \textsl{ROSAT} (Eckert et al. 2013a).}
\end{center}
\end{figure*}

\clearpage
\begin{figure*}
\begin{center}
\epsscale{1.15}\plottwo{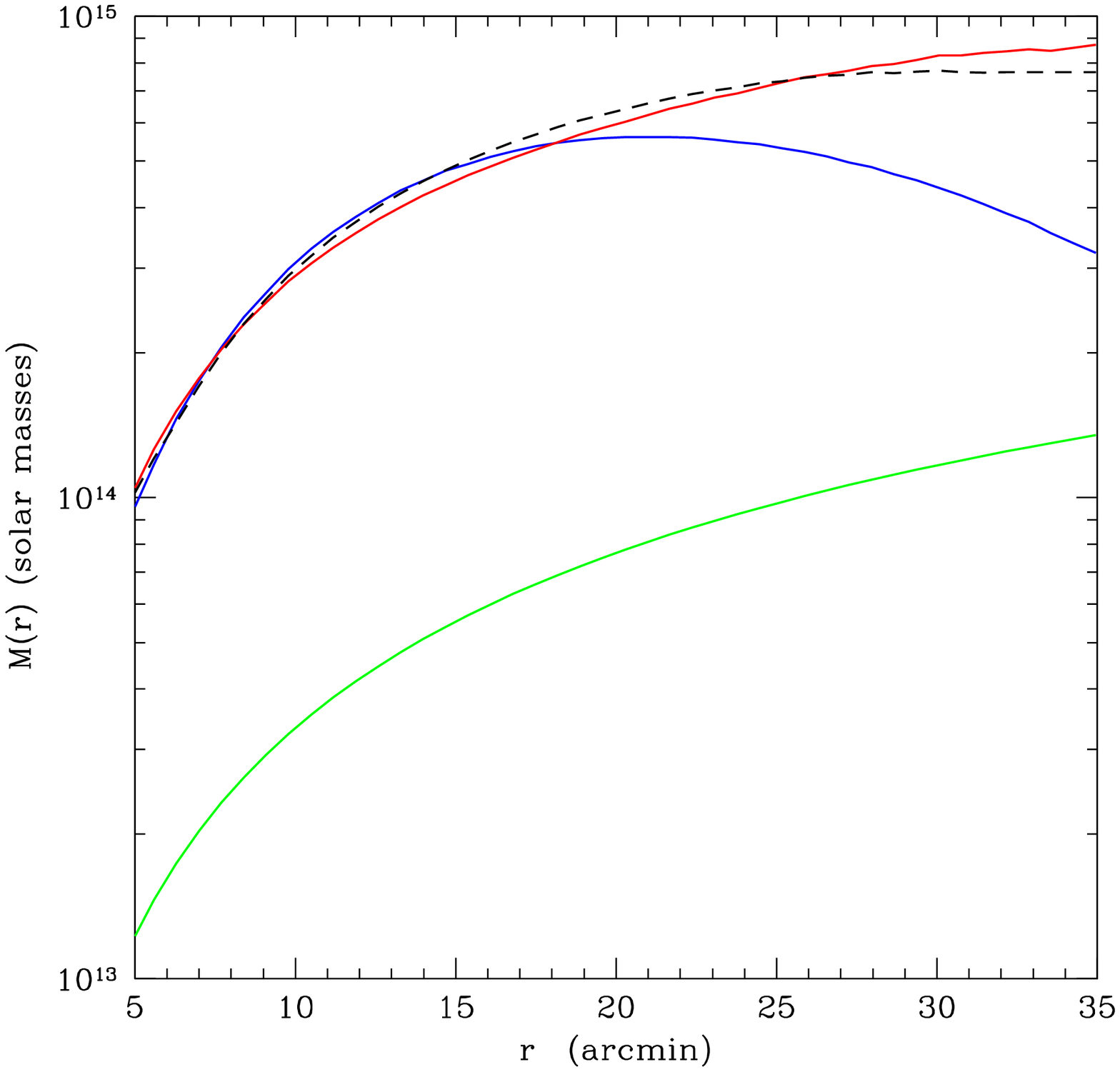}{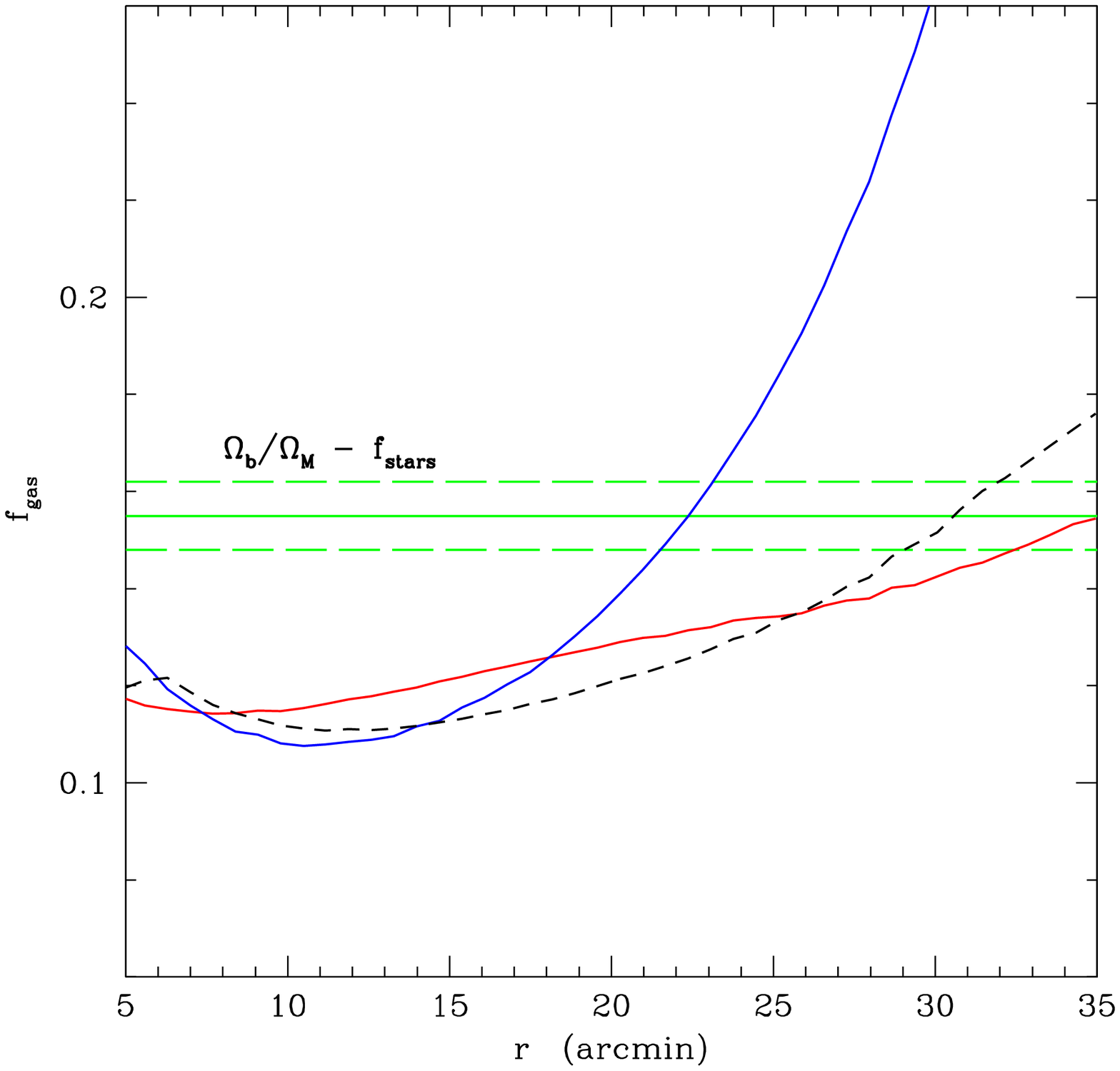}\caption{Abell
1795 - Left panel: Blue line is the total X-ray cluster mass obtained with
$\delta_R$ = 0; red and dashed black lines with $\delta_R$ = 1.3 and
$\delta_R$ = 1.1, respectively, and $l$ = 0.5; green line is the gas mass
obtained by the gas density of \textsl{ROSAT} (Eckert et al. 2013). Right
panel: Gas mass fraction derived from the above mass profiles; blue line is
with $\delta_R$ = 0; red and dashed black lines are with the above values of
$\delta_R$ and $l$; green lines are the difference of the cosmic baryon
fraction and the fraction of baryons in stars and galaxies,
$\Omega_b/\Omega_M -f_{stars} = 0.155 \pm 0.007$ (Komatsu et al. 2011;
Gonzalez et al. 2007).}
\end{center}
\end{figure*}

\clearpage
\begin{figure*}
\begin{center}
\epsscale{1.15}\plottwo{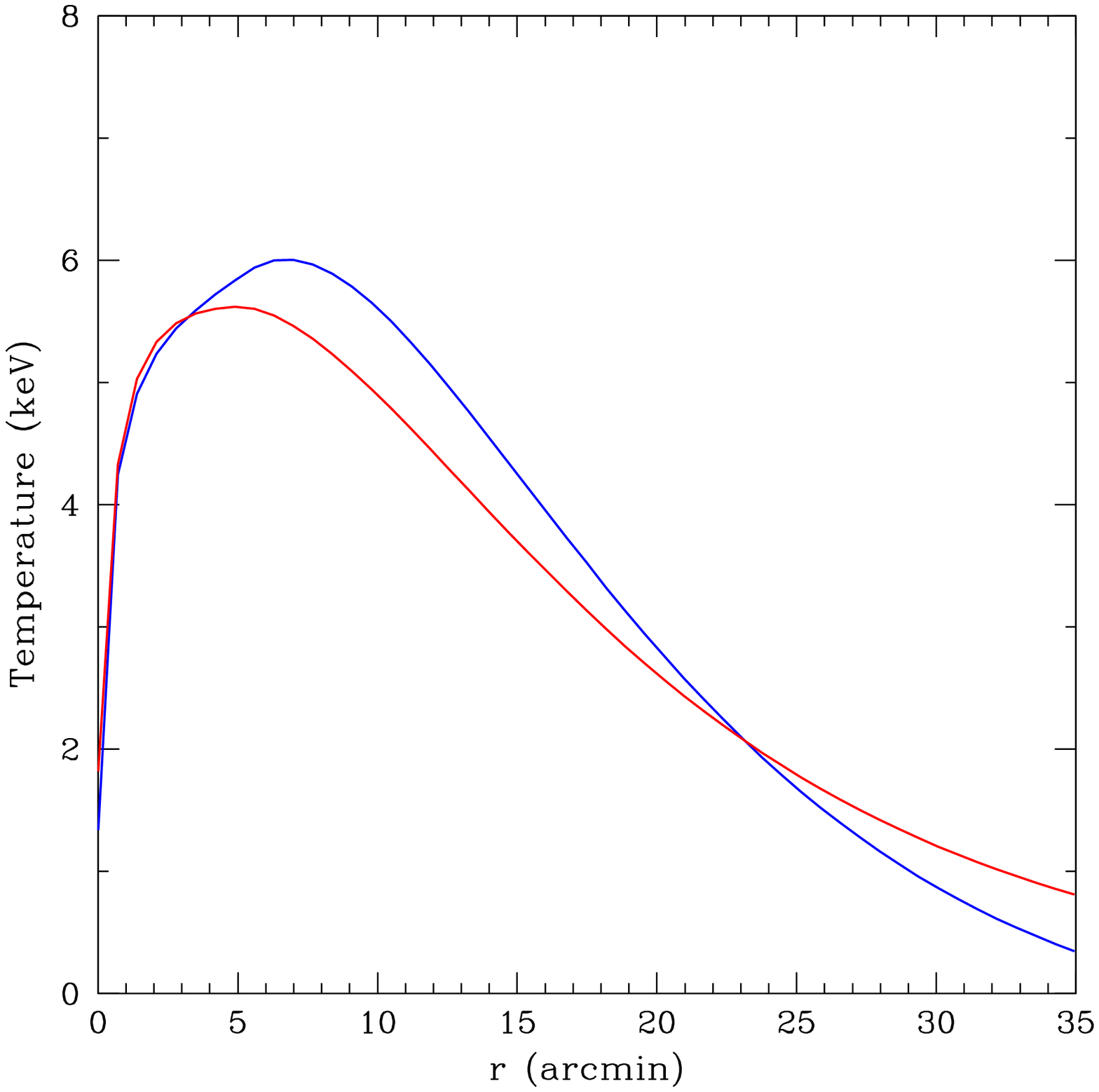}{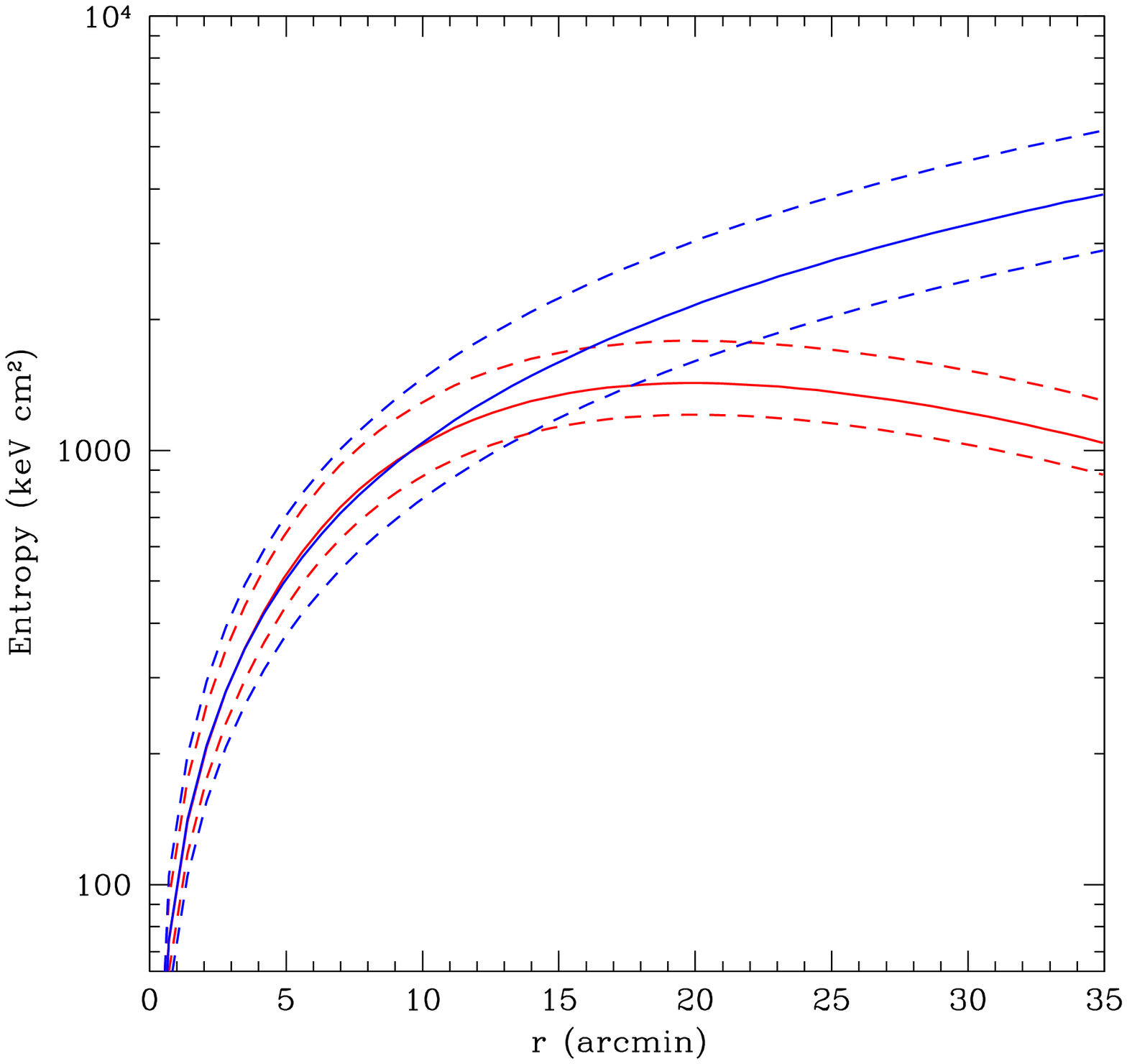}\caption{Abell
1795 - Left panel: Radial temperature profiles. Blue line is the radial
temperature obtained by the SM fit with $\delta_R$ = 0 to the projected
profiles of Fig.~1; red line is from the SM fit with $\delta_R$ = 1.3 and $l$
= 0.5. Both these profiles are obtained with entropy flattening. Right
panel: SM entropy profiles. Blue line is derived with the deprojected
temperature profile obtained by the SM fit to the projected temperature
profile with a power law increase of the entropy (see dashed green line of
Fig.~1); red line from the deprojected temperature derived by the SM fit to
the projected temperature profile (see red line of Fig.~1) with $\delta_R$ =
1.3, $l$ = 0.5 and entropy flattening. Dashed lines are the 68\% confidence
intervals.}
\end{center}
\end{figure*}

\clearpage
\begin{figure*}
\begin{center}
\epsscale{1.15}\plottwo{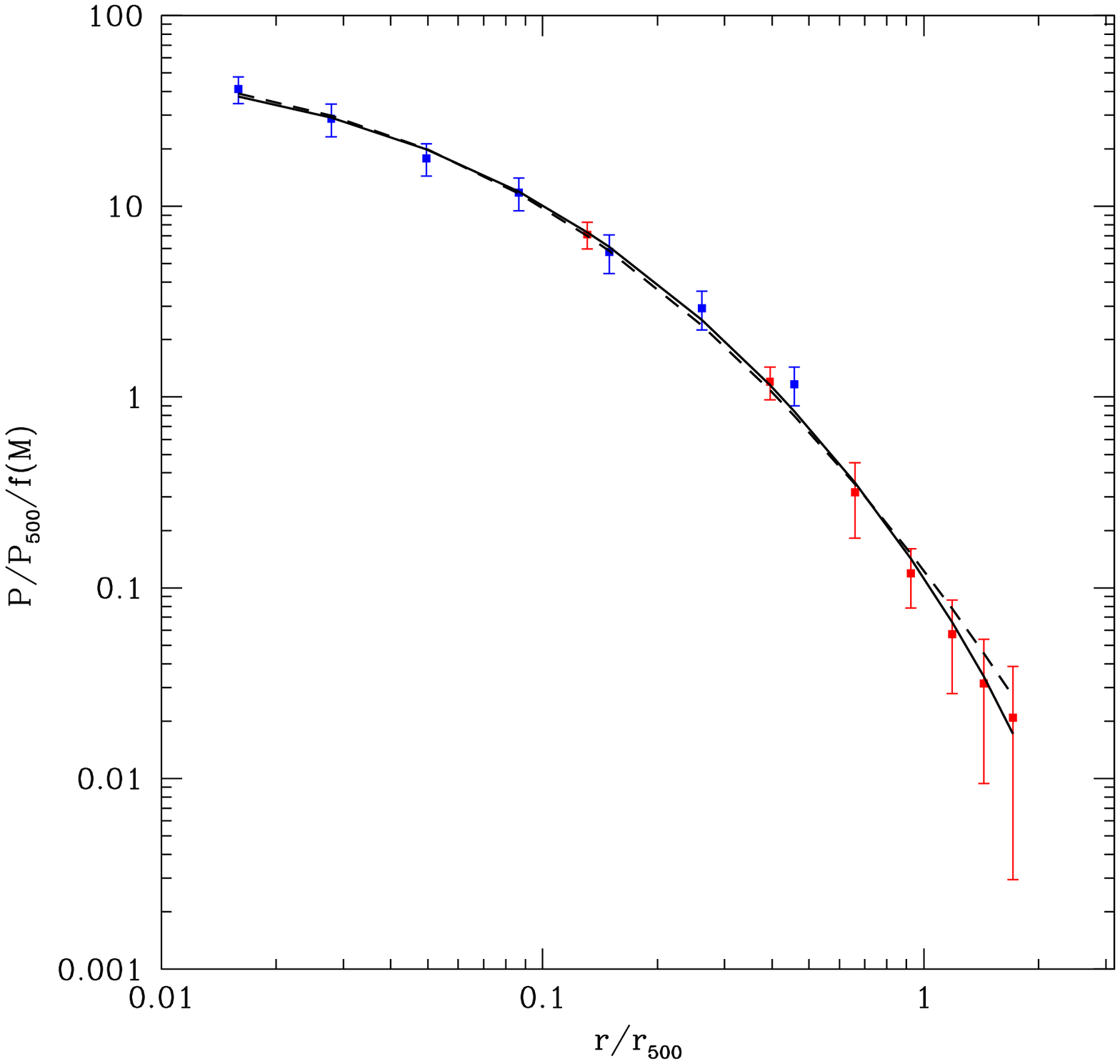}{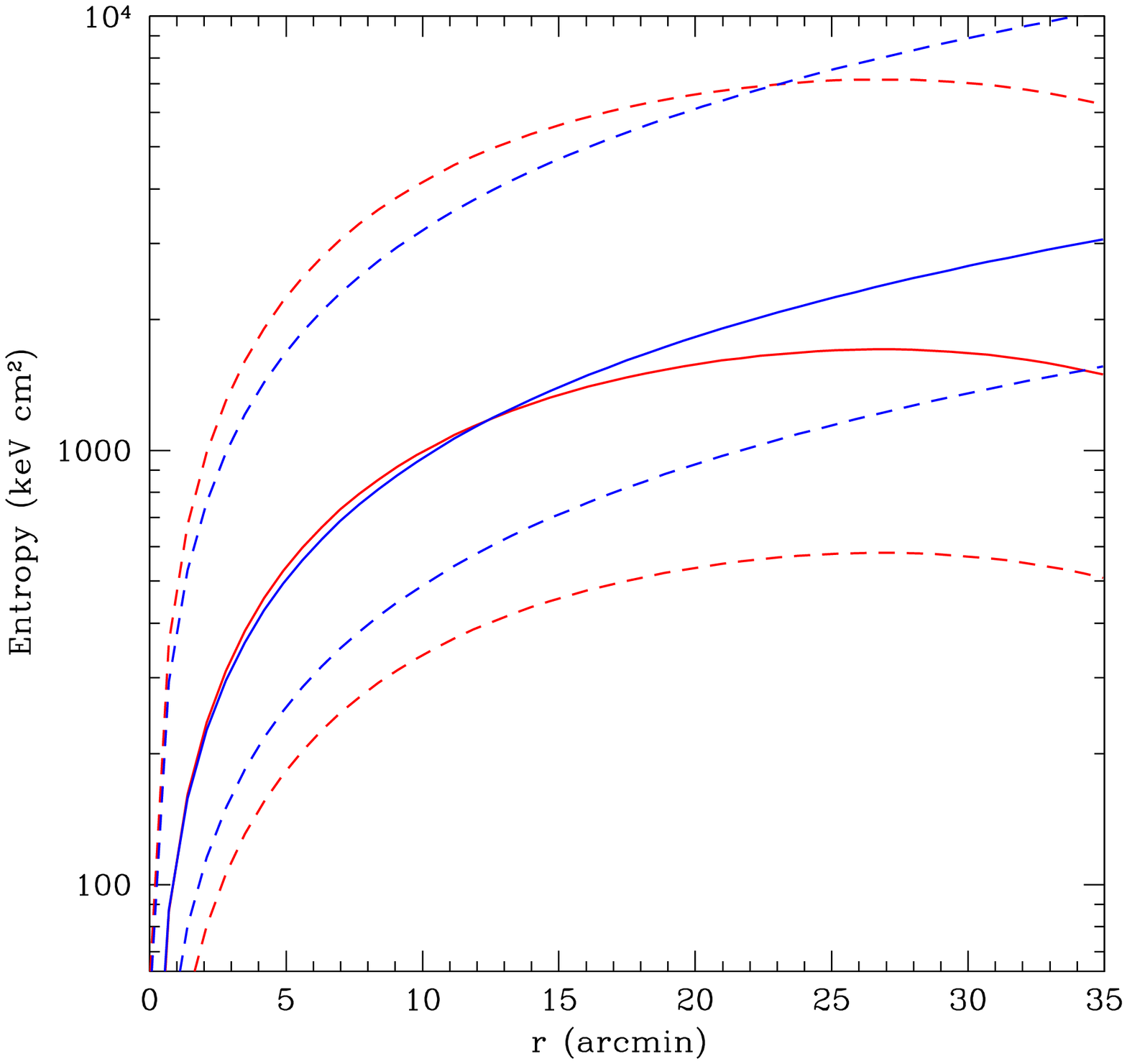}\caption{Abell
1795 - Left panel: Pressure profiles using \textsl{XMM-Newton} (blue points)
and \textsl{Planck} (red points) data (\textsl{Planck} Collaboration
2013a); dashed line is the SM fit with an entropy power law increase while
the continuous line is with a flattening of the entropy at $r > r_b$. Right
panel: Entropy profiles obtained by the above pressure profiles ($k =
P/n_e^{5/3}$); blue line with a power law increase of the entropy and red
line with an entropy flattening. The dashed curves are the 68\% confidence
intervals.}
\end{center}
\end{figure*}

\clearpage
\begin{figure*}
\begin{center}
\epsscale{1.15}\plottwo{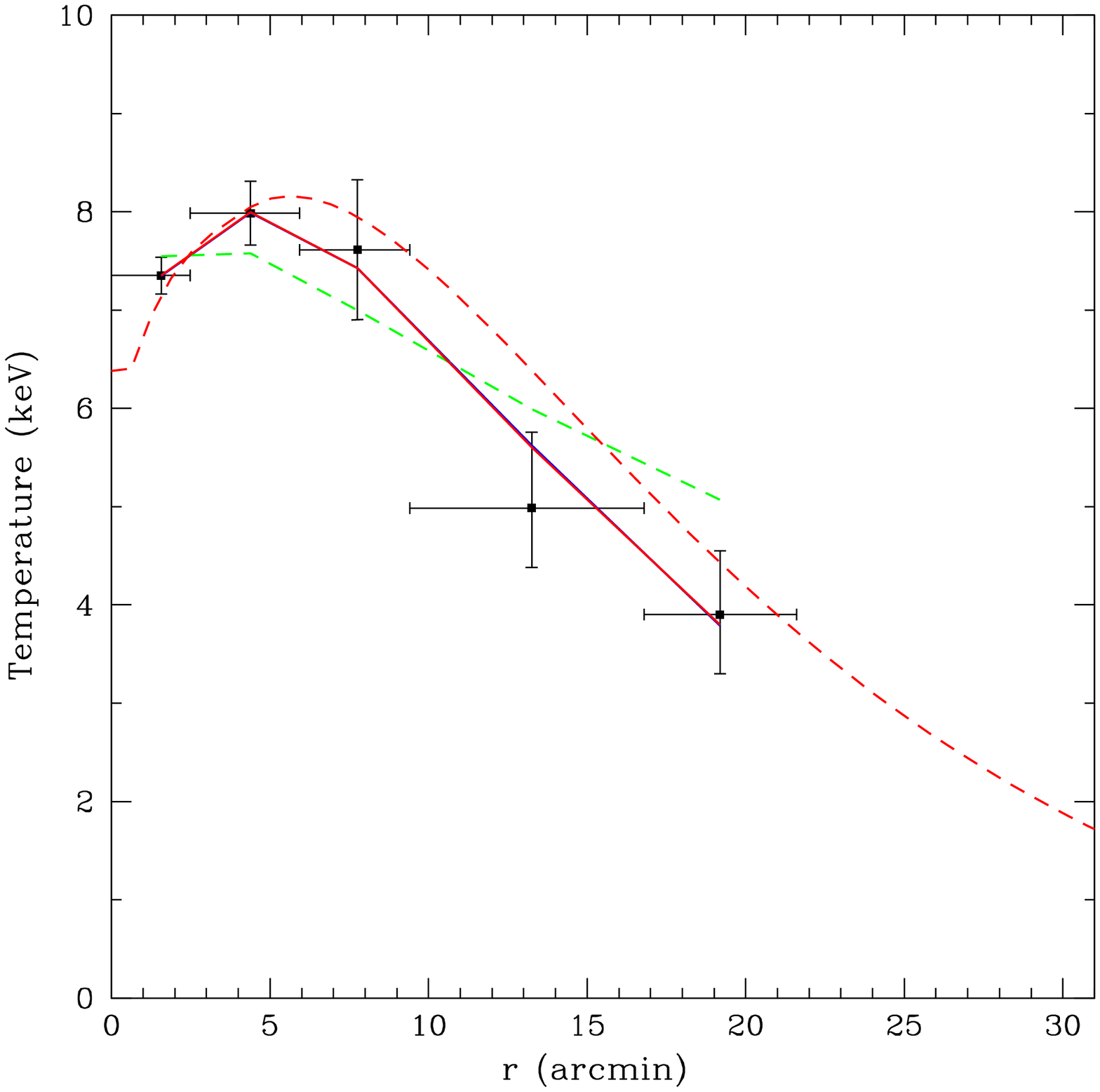}{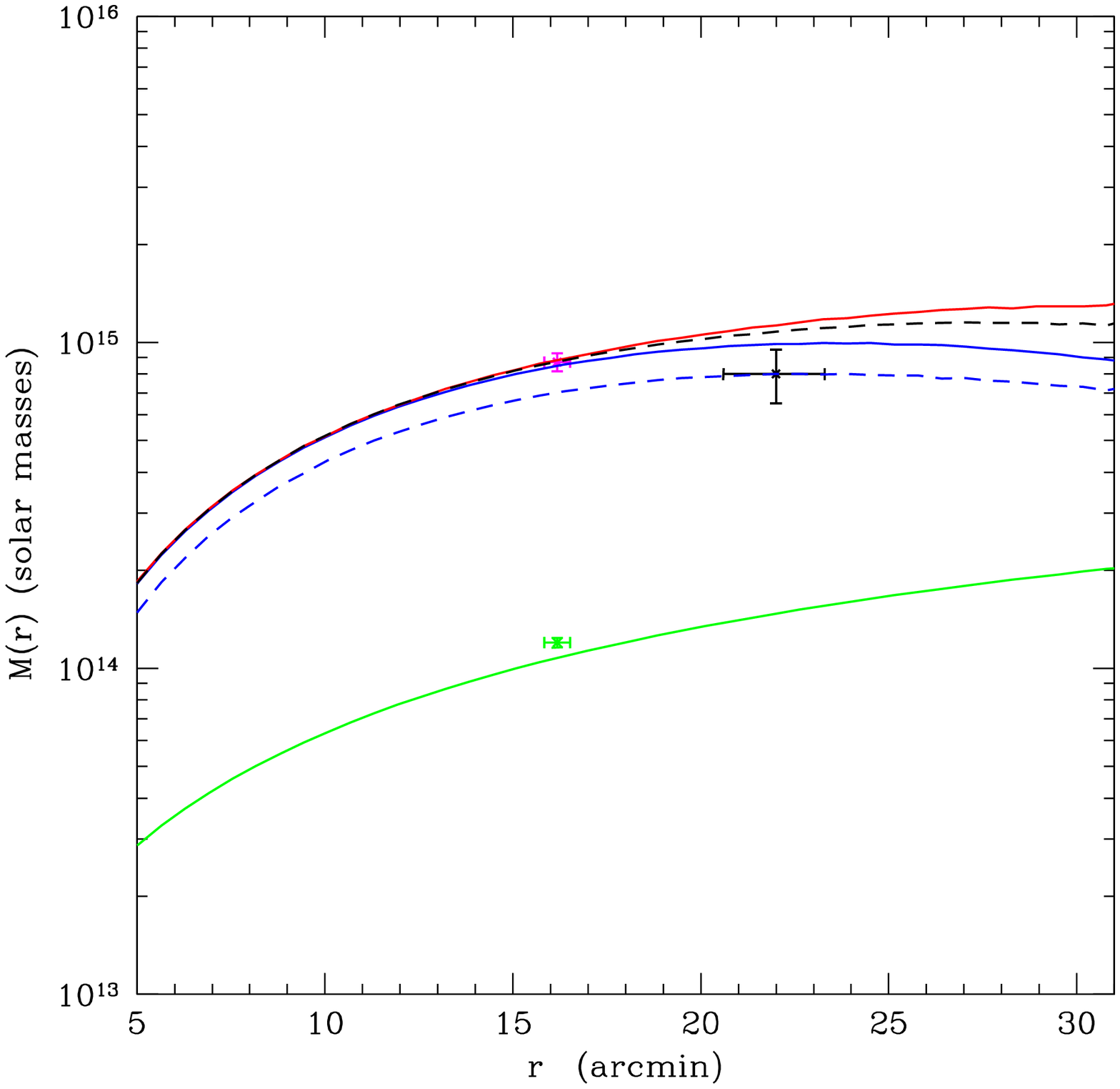}\caption{Abell
2029 - Left panel: Azimuthally averaged (excluding the north and the SE)
temperature profile observed by \textsl{Suzaku} (Walker et al. 2012a). Blue
line is the SM fit with entropy flattening and $\delta_R$ = 0; red line with
$\delta_R$ = 0.5, $l$ = 0.5 (the blue and red lines are practically
coincident); dashed green line is the SM fit with a power law for the entropy
run; dashed red line is the deprojected temperature profile. Right panel:
Blue line is the total X-ray cluster mass with $\delta_R$ = 0, red line with
$\delta_R$ = 0.5 and $l$ =0.5, dashed black line with $\delta_R$ = 0.3 and
$l$ = 0.5; green line is the gas mass profile obtained by \textsl{ROSAT}
observations (Eckert et al. 2013). Dashed blue line is the mass profile
obtained using the SM deprojected temperature profile (left panel) and the
\textsl{Suzaku} density profile (Walker et al. 2012a). The black point is
their value derived at $\,r_{200}$. The magenta point is the X-ray mass at
$r_{500}$ derived by Gonzalez et al. (2013) from the \textsl{XMM-Newton}
data; the green point is the gas mass}
\end{center}
\end{figure*}

\clearpage
\begin{figure*}
\begin{center}
\epsscale{1.15}\plottwo{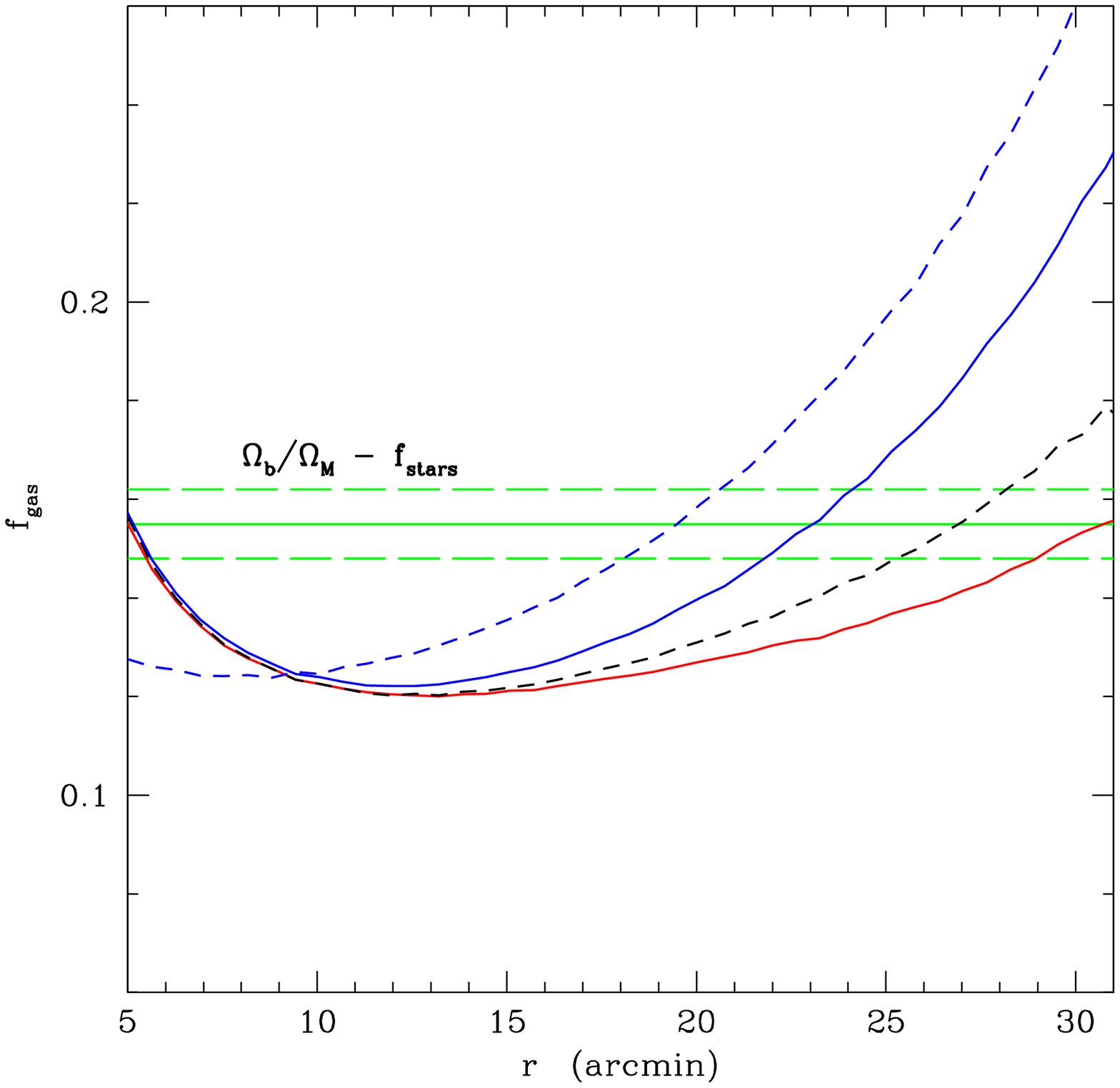}{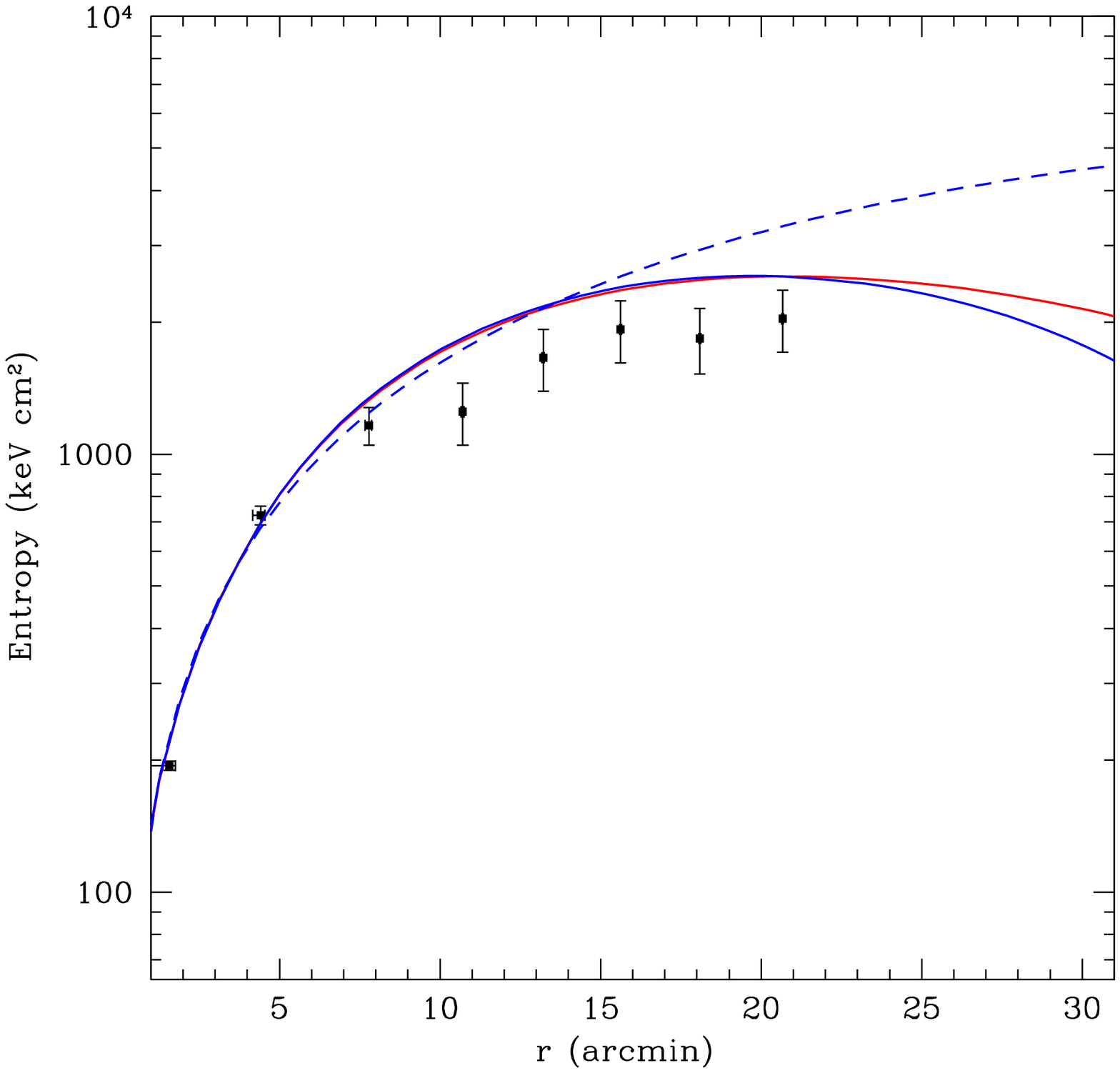}\caption{Abell
2029 - Left panel: Gas mass fraction. Blue, red and dashed black lines refer
to the mass profile with the same color of Fig.~5. Dashed blue line is
obtained with the gas density profile of \textsl{Suzaku} (Walker et al.
2012a). Right panel: Entropy profiles derived with the deprojected
temperature profiles obtained by the SM fits to the temperature profile of
Fig.~5; blue line with $\delta$ = 0, and red line with $\delta_R$ = 0.5, $l$
= 0.5. Both with entropy flattening. Dashed blue line from the fit to the
temperature profile with a power law increase of the entropy (dashed green
line of Fig.~5). The points are taken by Fig.~6 of Walker et al. (2012a) that
consider the \textsl{Suzaku} gas density profile.}
\end{center}
\end{figure*}

\clearpage
\begin{figure*}
\begin{center}
\epsscale{1.15}\plottwo{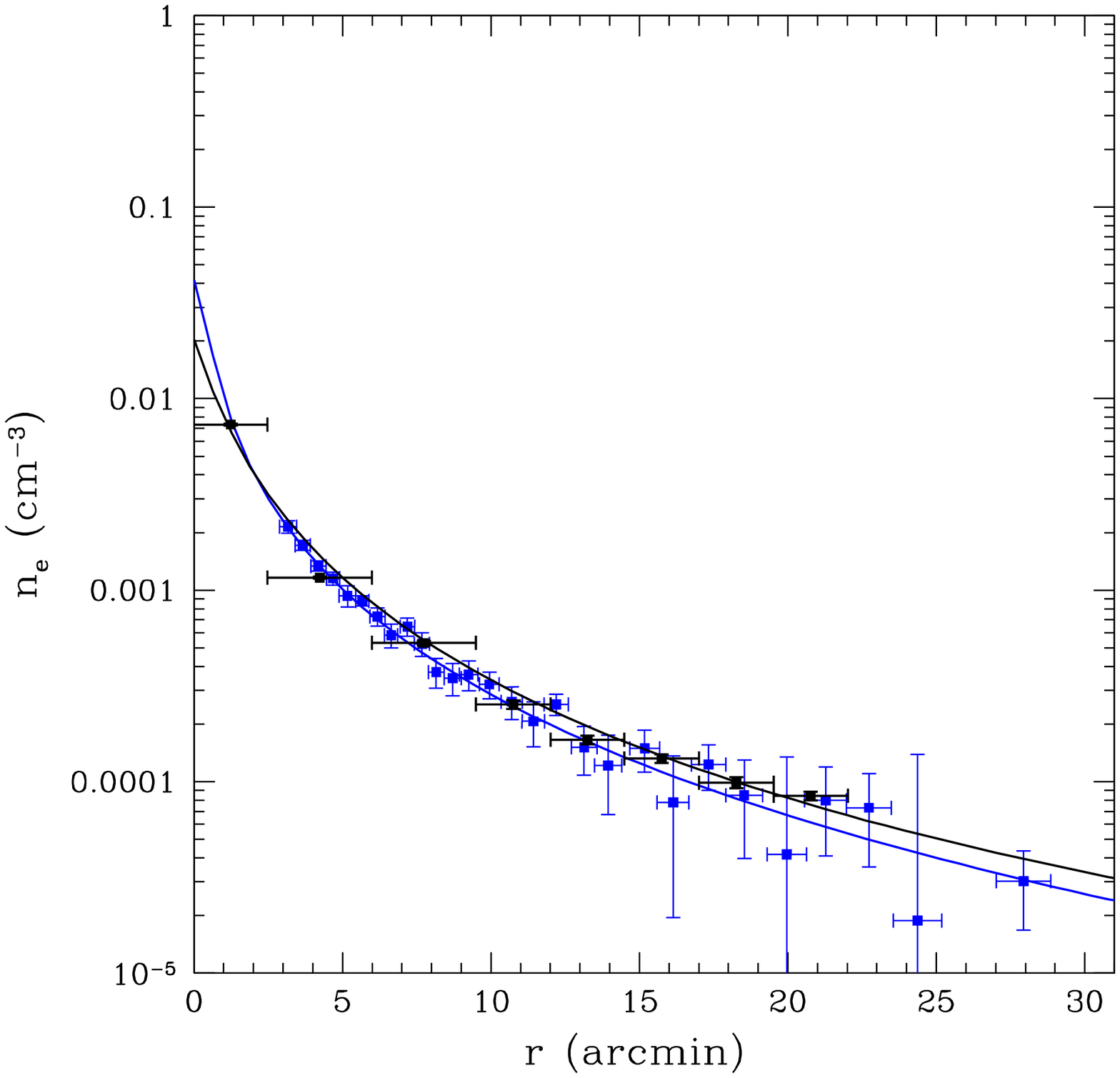}{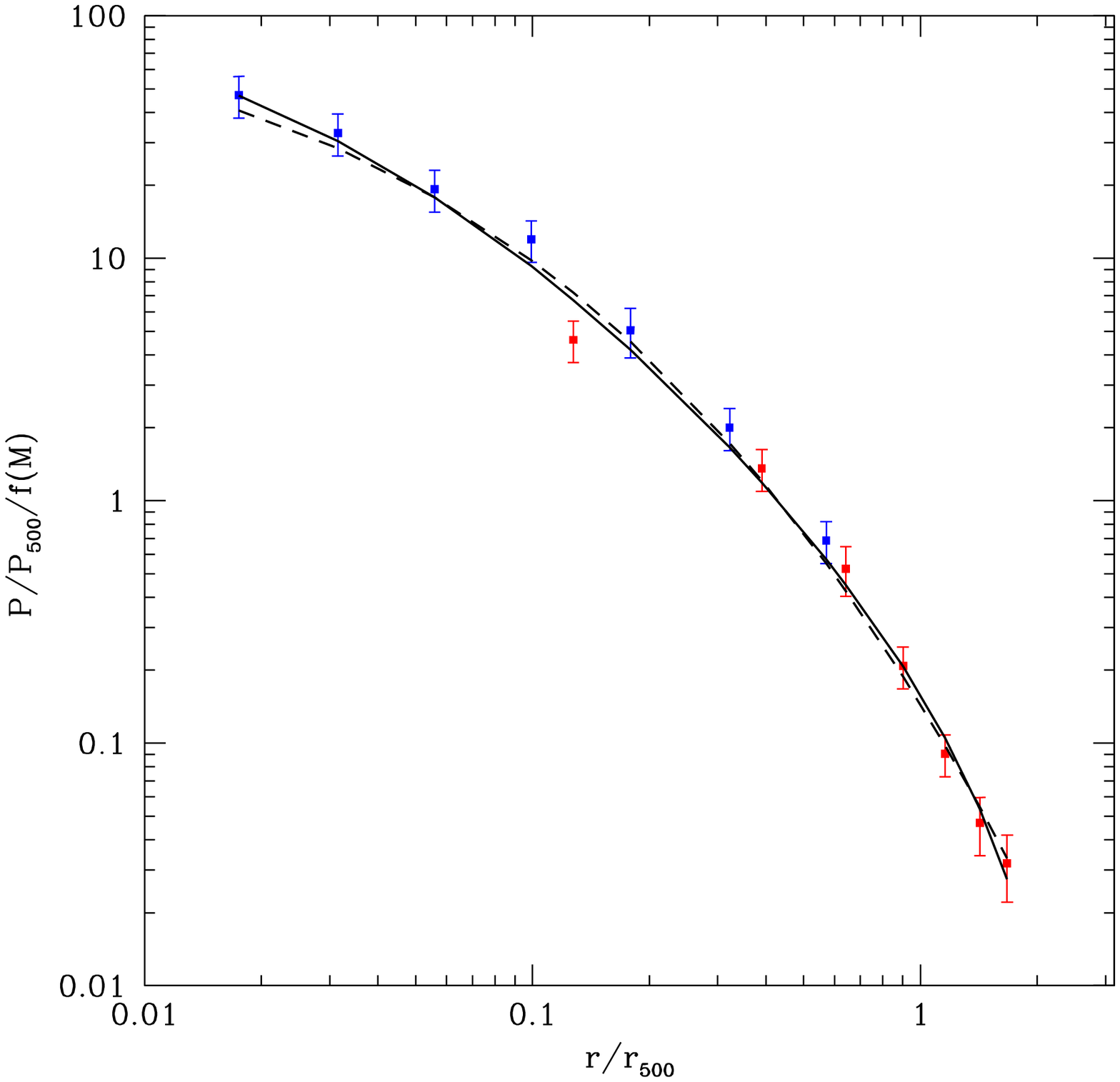}\caption{Abell
2029 - Left panel: Gas density. Blue line is the SM fit to the blue electron
density points of \textsl{ROSAT} (Eckert et al. 2013); black line is the SM
fit to the black electron density points of \textsl{Suzaku} (Walker et al.
2012a). Right panel: Pressure profiles using \textsl{XMM-Newton} (blue
points) and \textsl{Planck} (red points) data (\textsl{Planck}
Collaboration 2013a); dashed line is the SM fit with an entropy power law
increase while the continuous line is with a flattening of the entropy at $r
> r_b$.}
\end{center}
\end{figure*}

\clearpage
\begin{figure*}
\begin{center}
\epsscale{1.15}\plottwo{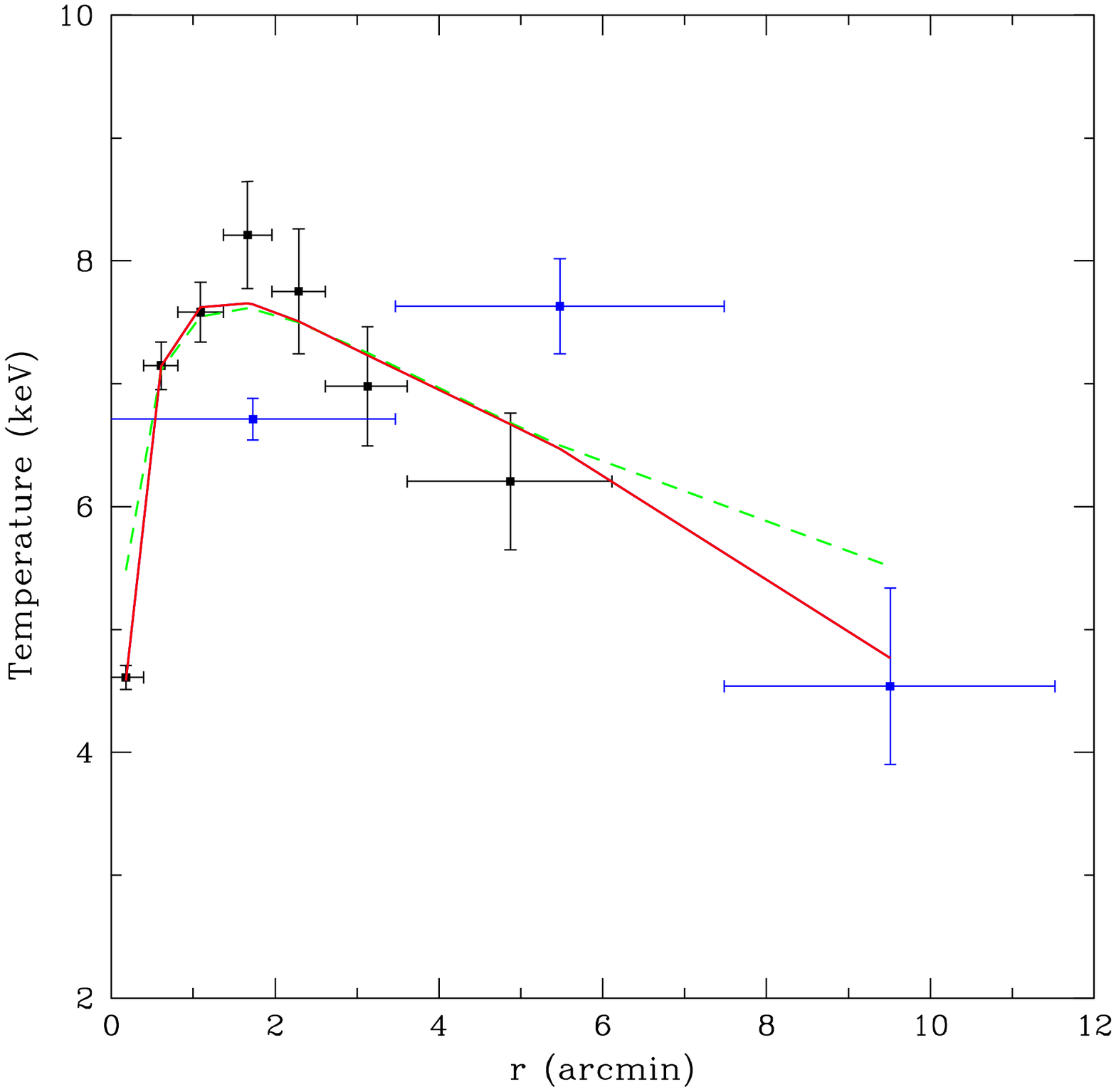}{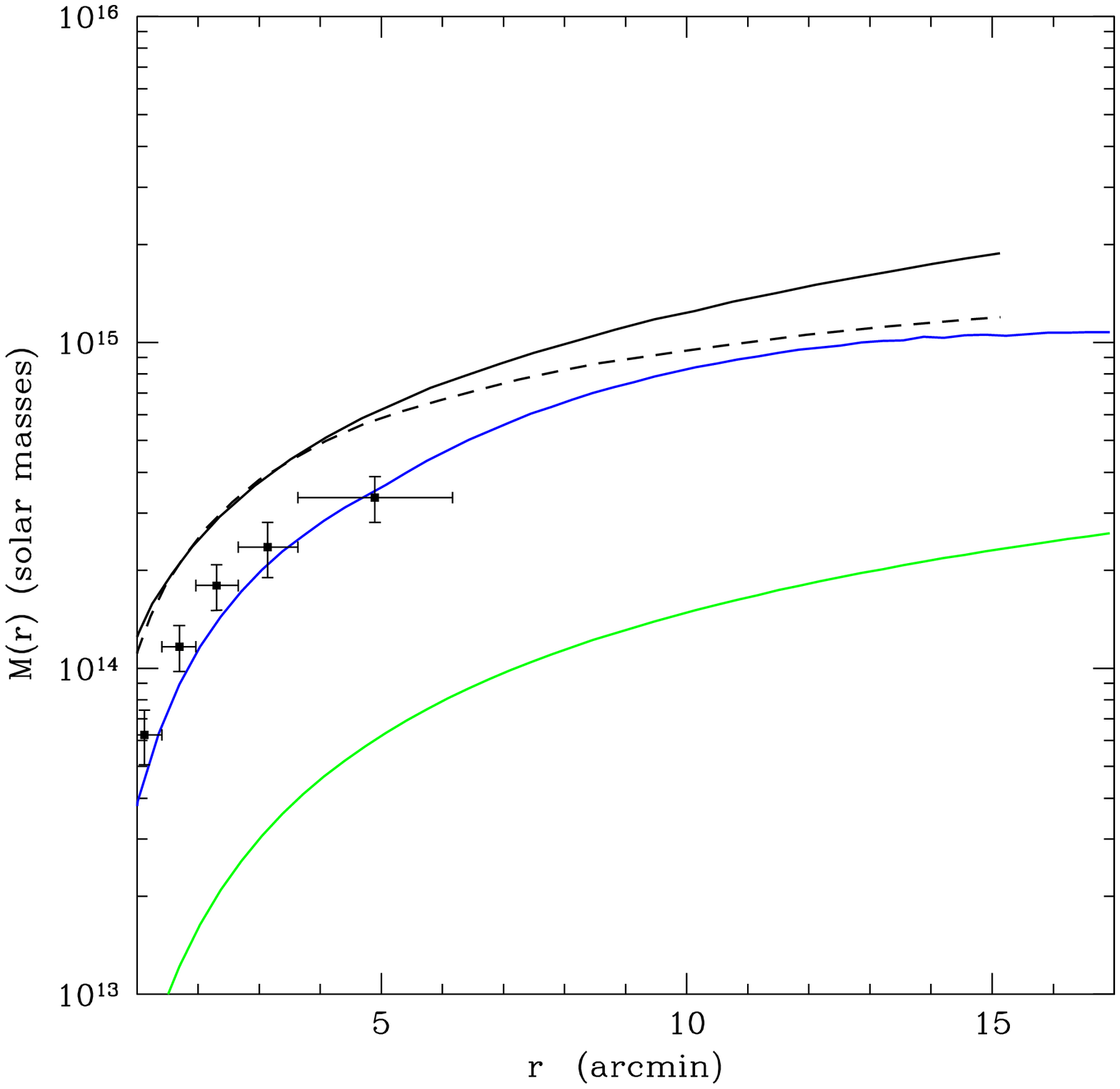}\caption{Abell
2204 - Left panel: Temperature profiles. Blue points from \textsl{Suzaku}
(Reiprich et al. 2009), black points from \textsl{XMM-Newton} (Zhang et al.
2008). Blue and red lines (practically coincident) are the SM fits with
entropy flattening and with $\delta_R$ = 0 and $\delta_R$ = 0.15 ($l$ = 0.5),
respectively. Dashed green line is with a power law entropy profile. Right
panel: Total cluster mass. Blue line is derived with the deprojected
temperature profile obtained by the SM fit of the same color to the projected
temperature profile in the left panel; points are from the \textsl{XMM-Newton}
analysis within $r_{500}$ (Zhang et al. 2008). Dashed and continuous black
lines ar the best fit NFW and SIS models to the weak lensing data (Clowe \&
Schneider 2002); the green line is the gas mass profile obtained by the gas
density of \textsl{ROSAT} (Eckert et al. 2013).}
\end{center}
\end{figure*}

\clearpage
\begin{figure*}
\begin{center}
\epsscale{1.15}\plottwo{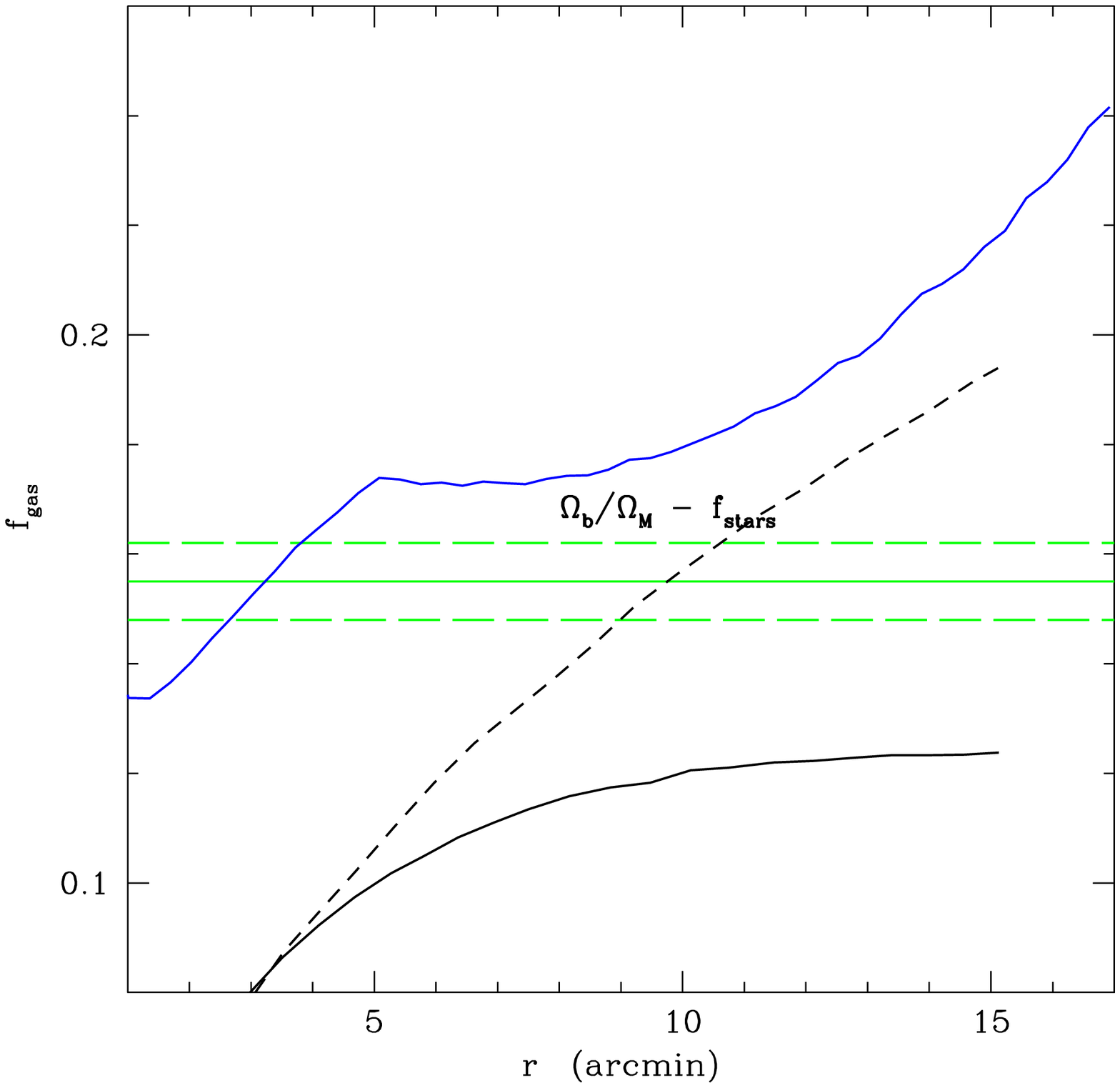}{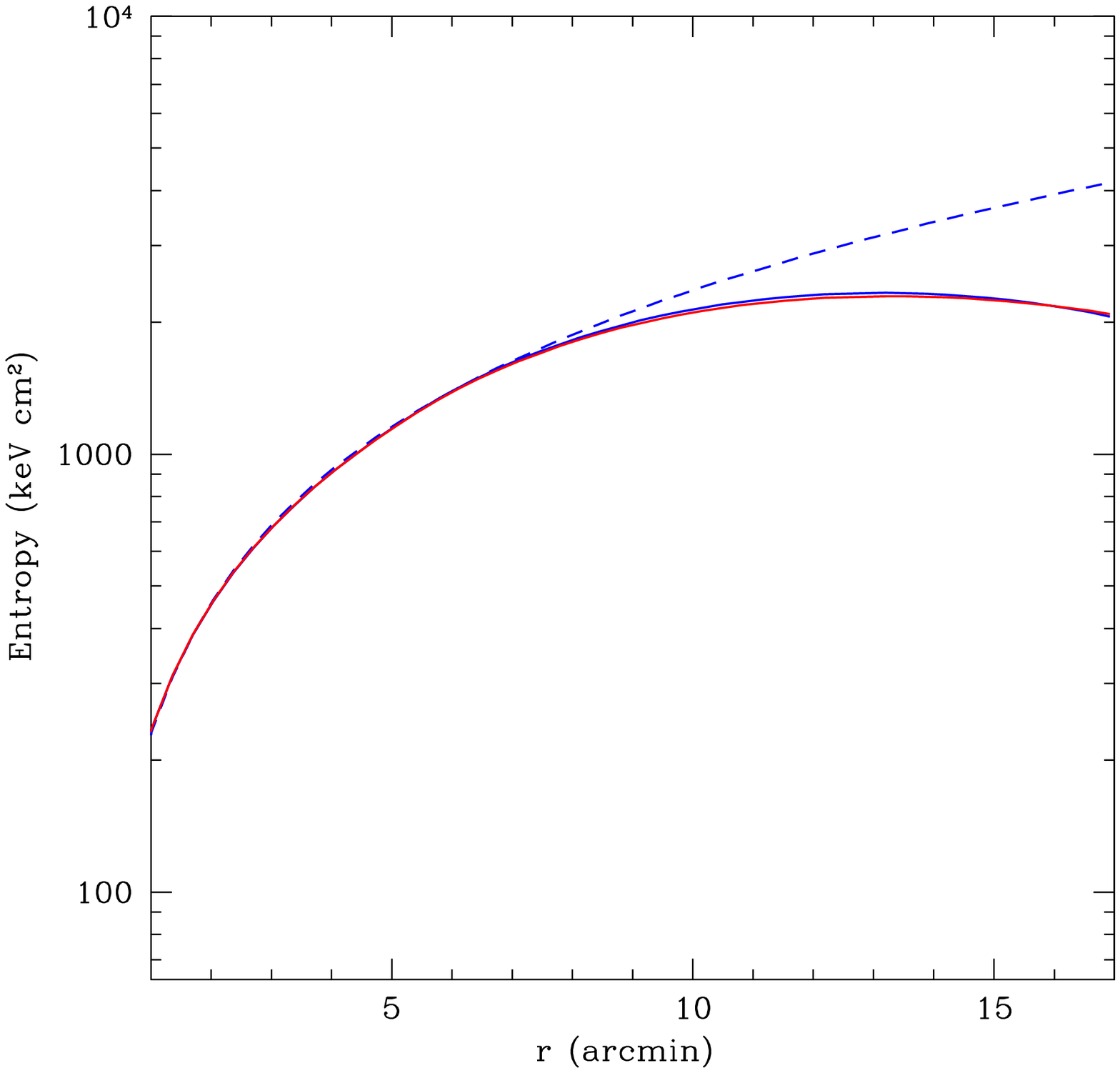}\caption{Abell
2204 - Left panel: Gas mass fraction. Blue line is derived by the cluster mass
profile of the same color (see Fig.~8). The dashed and continuous lines are
derived with the NFW and SIS models, respectively. Right panel: Entropy
profiles. Blue and red lines derived with the deprojected temperature
profiles obtained by the fits of the same colors to the projected temperature
profile of Fig.~8. Dashed blue line is with the deprojected temperature
profile obtained by the SM fit with a power law increase of the entropy
(green dashed line of Fig.~8).}
\end{center}
\end{figure*}

\clearpage
\begin{figure*}
\epsscale{0.8}\plotone{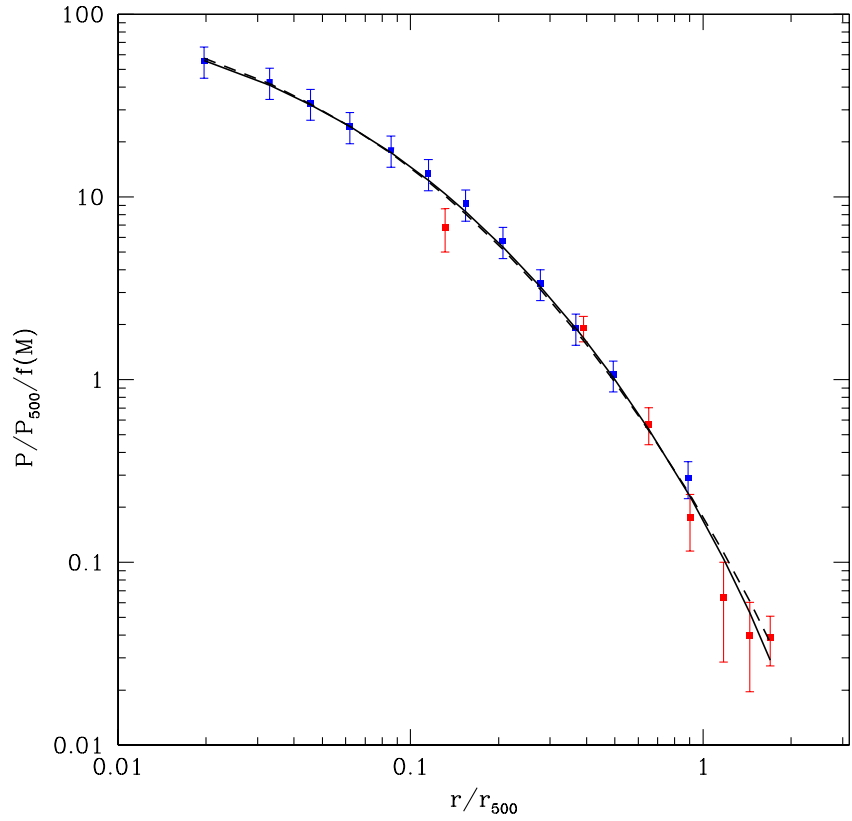}\caption{Abell 2204 - Pressure profiles
using \textsl{XMM-Newton} (blue points) and \textsl{Planck} (red points) data
(\textsl{Planck} Collaboration 2013a); dashed line is the SM fit with an
entropy power law increase while the continuous line is with a flattening of
the entropy at $r > r_b$.}
\end{figure*}

\clearpage
\begin{figure*}
\begin{center}
\epsscale{1.15}\plottwo{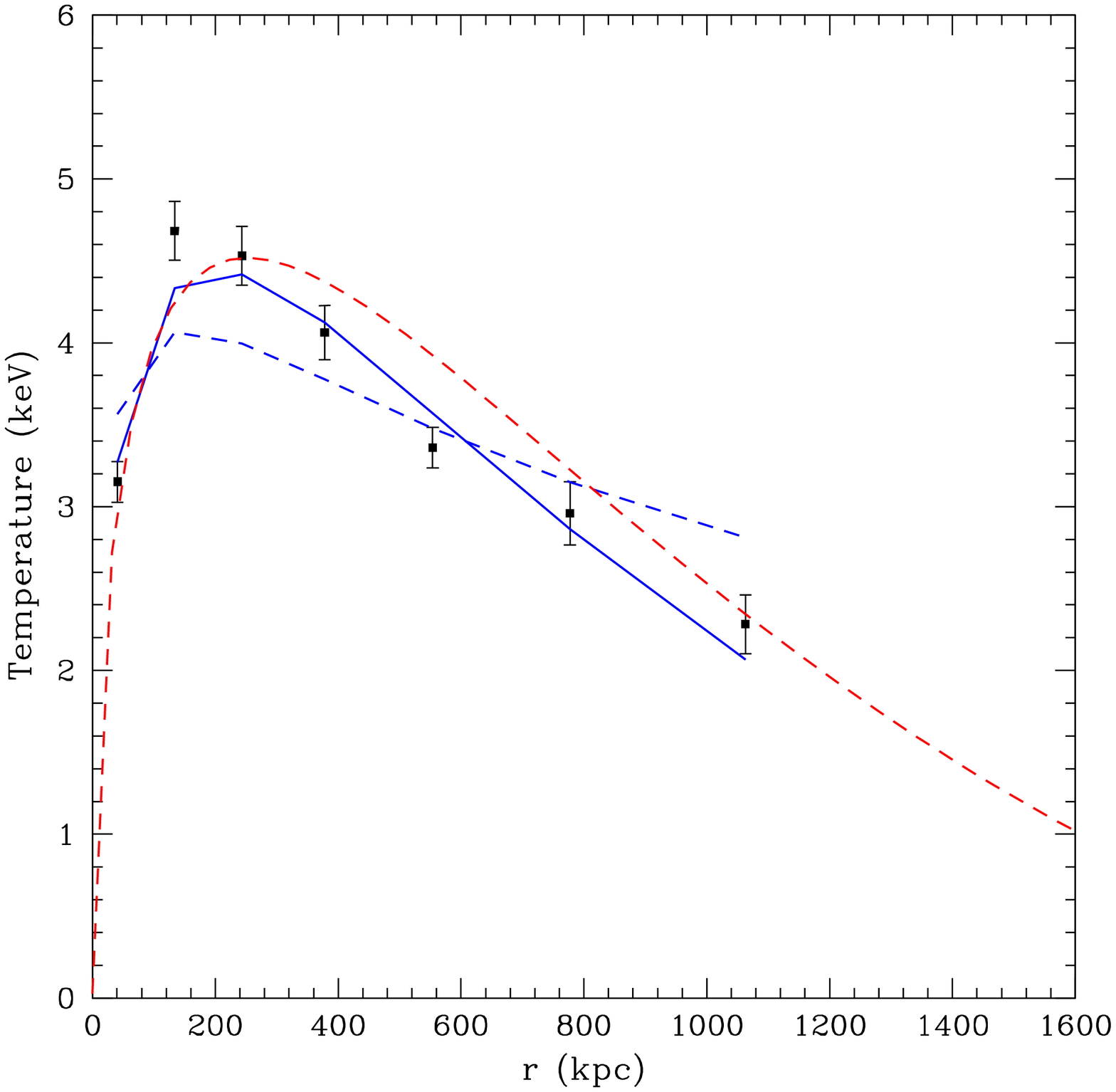}{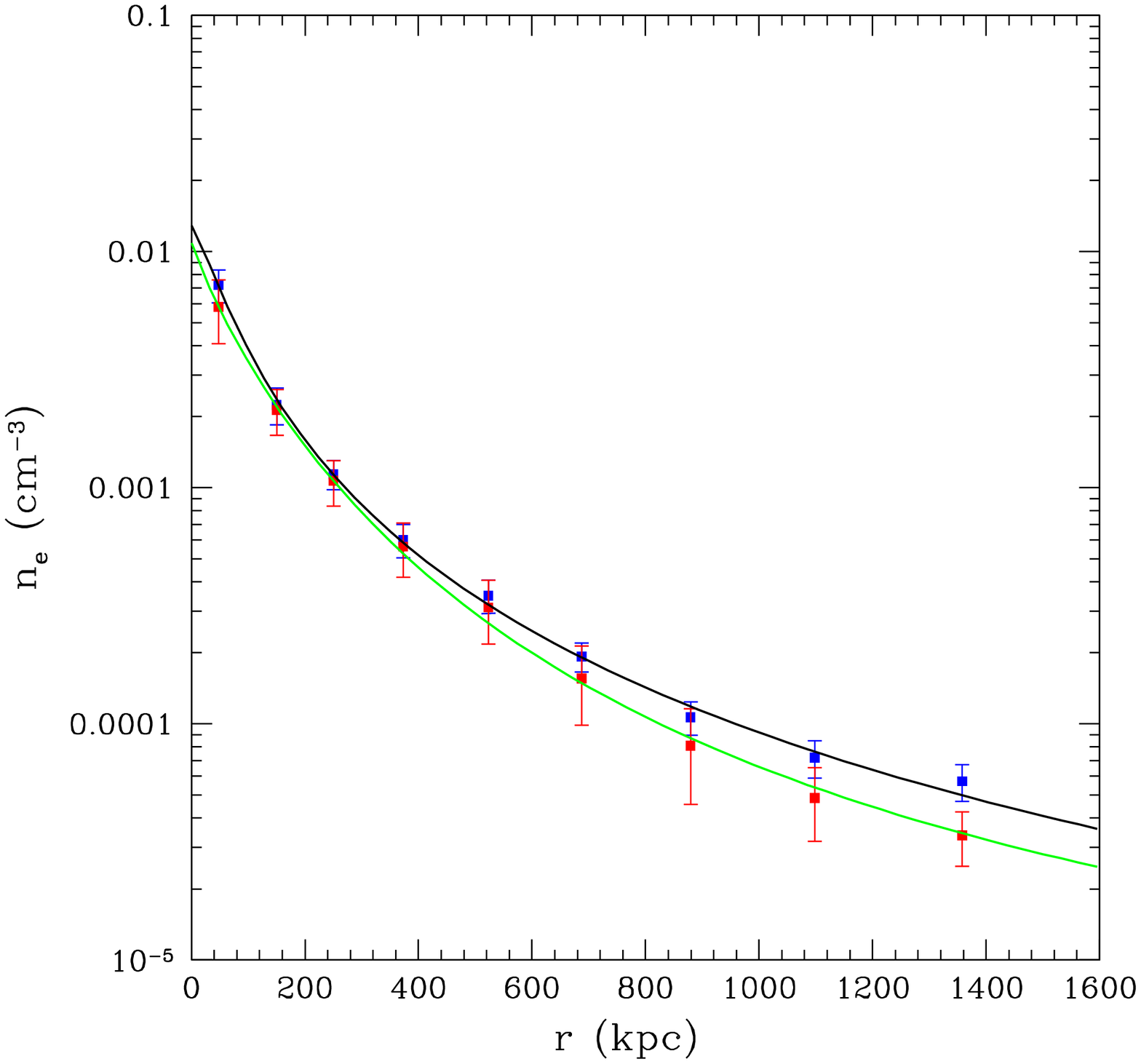}\caption{Abell
133 - Left panel: Temperature profiles. Points from \textsl{Chandra} (Morandi
\& Cui 2013); continuous line is the SM fit with entropy flattening. Blue
dashed line is with a power law increase for the entropy; dashed red line is
the deprojected temperature. Right panel: Gas density profiles. Blue points
from the \textsl{Chandra} analysis; red points are obtained when the gas
clumping effect is taken into account (Morandi \& Cui 2013). Black and green
lines are the SM fits, respectively.}
\end{center}
\end{figure*}

\clearpage
\begin{figure*}
\begin{center}
\epsscale{1.15}\plottwo{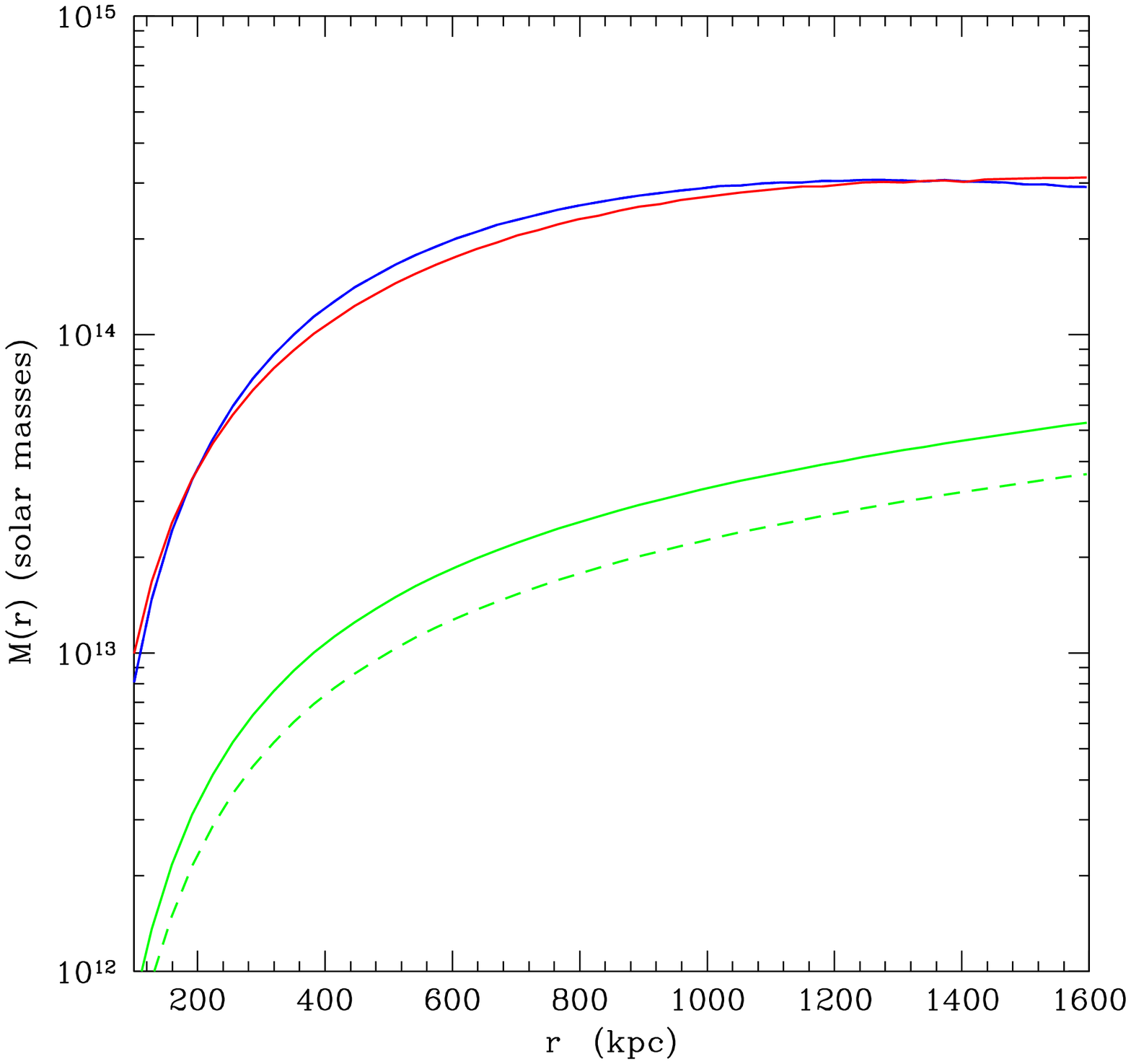}{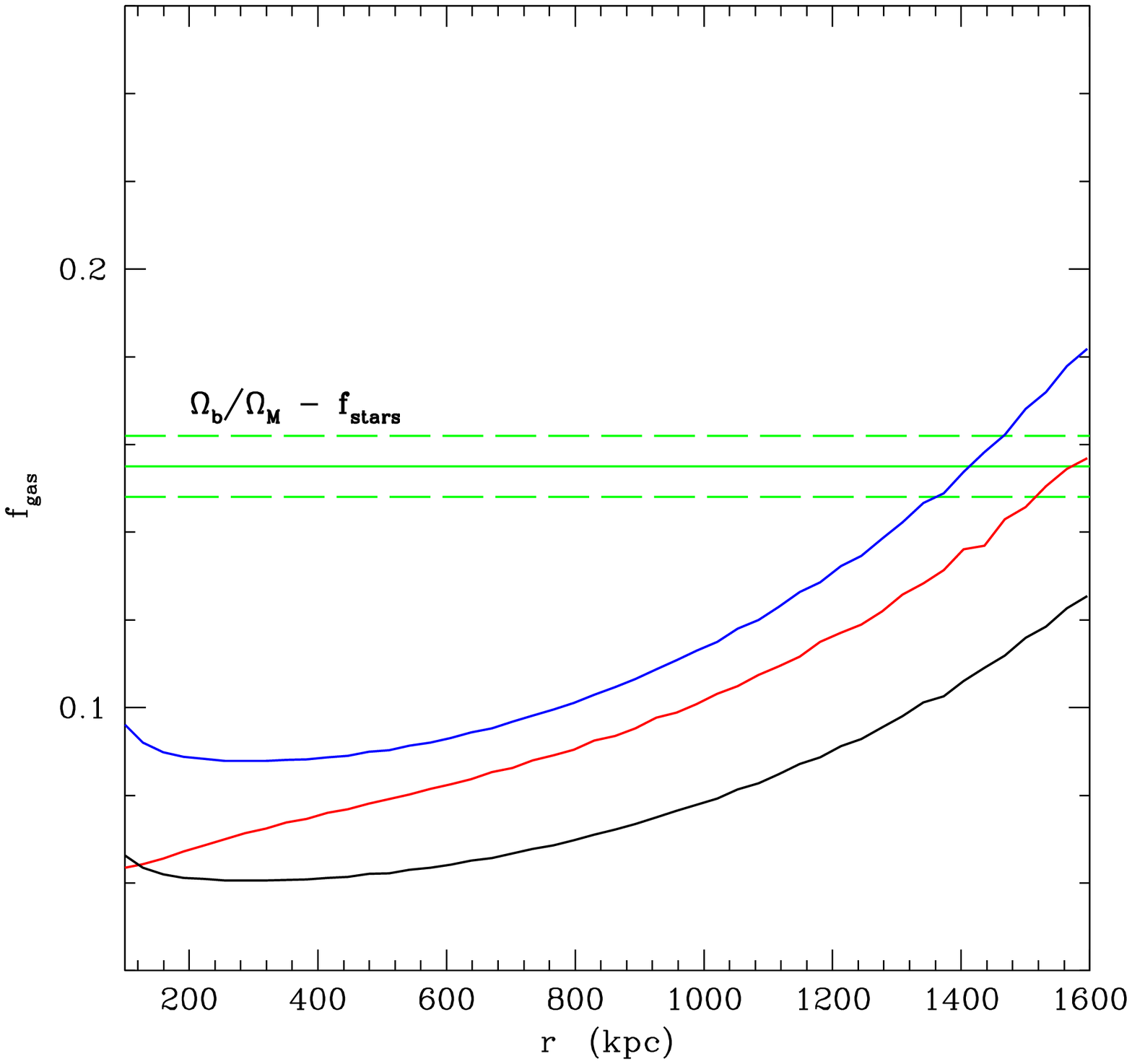}\caption{Abell
133 - Left panel: total X-ray cluster mass. Blue line is derived with the
deprojected temperature profile of Fig.~11 (dashed red line); red line is
with the deprojected temperature profile obtained with $\delta_R$ = 0.05 and
$l$ = 0.5. Continuous green line is the gas mass profile derived with the
density profile of Fig.~11 (black line); dashed green line is the gas mass
derived with the density profile (green line of Fig.~11) that considers the
gas clumping effect . Right panel: Gas mass fraction. Blue line is derived
with the total cluster mass given by the blue line and continuous green line
for the gas mass (see left panel); red line is with $\delta_R$ = 0.05, $l$ =
0.5; black line is derived with the gas mass given by the green dashed line
of the left panel.}
\end{center}
\end{figure*}

\clearpage
\begin{figure*}
\epsscale{0.8}\plotone{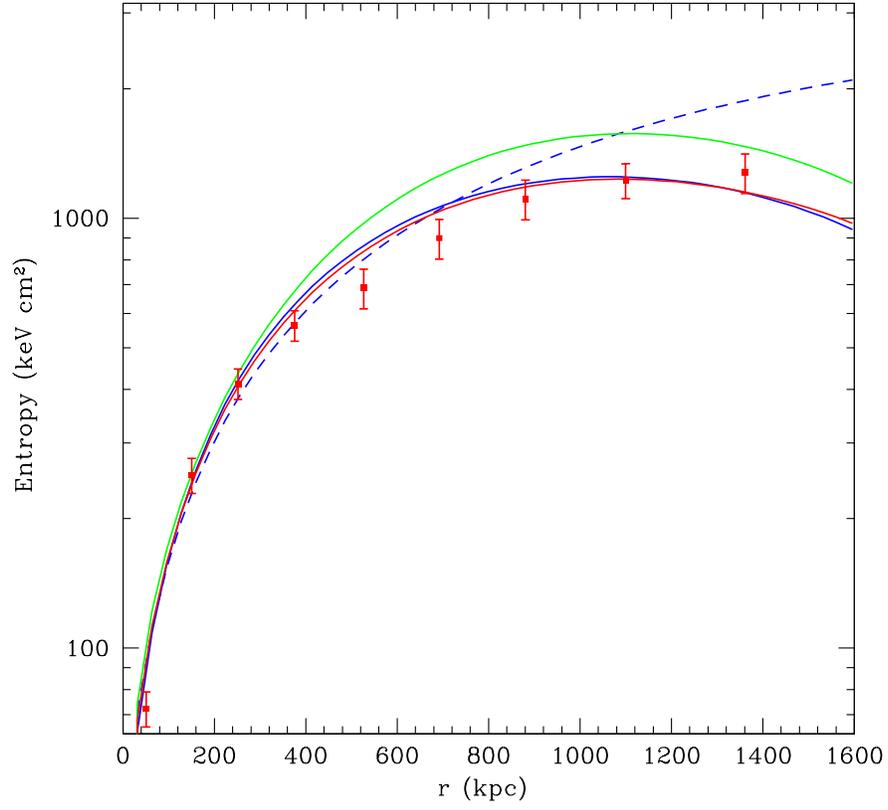}\caption{Abell 133 - Entropy
profiles. Blue and red lines are derived from the temperature profiles with
$\delta_R$ = 0 and $\delta_R$ = 0.05, respectively. Dashed blue line is
obtained with the temperature profile (dashed blue line of Fig.~11) derived
with a power law increase for the entropy; green line is with the gas density
that considers the gas clumping effects (green line of Fig.~11). Points are
from the \textsl{Chandra} analysis of Morandi \& Cui (2013) with the gas
density of Fig.~11 (black line).}
\end{figure*}

\end{document}